\documentclass[eqsecnum,amsmath,preprintnumbers,superscriptaddress,nofootinbib,aps,10pt]
{revtex4}
\pdfoutput=1
\usepackage{tikz}
\usetikzlibrary{calc,trees,positioning,arrows,chains,shapes.geometric,decorations.pathreplacing,decorations.pathmorphing,shapes,matrix,shapes.symbols}

\usepackage{bm, adjustbox}
\usepackage{MnSymbol, threeparttable}
\usepackage{float}[H]
\usepackage{amsmath}
\usepackage{dsfont}
\usepackage{graphicx}
\usepackage{feynmp}
\usepackage{hyperref}
\usepackage{bbold}
\usepackage{eufrak, comment, mathtools}
\usepackage{upgreek}
\usepackage{stmaryrd,scalerel}
\usepackage{mathtools}
\usepackage{enumitem}
\usepackage{color} 
\usepackage{etex}
\usepackage{morefloats}

\usepackage{scalerel,stackengine}
\stackMath
\newcommand\reallywidehat[1]{%
	\savestack{\tmpbox}{\stretchto{%
			\scaleto{%
				\scalerel*[\widthof{\ensuremath{#1}}]{\kern-.6pt\bigwedge\kern-.6pt}%
				{\rule[-\textheight/2]{1ex}{\textheight}}
			}{\textheight}%
		}{0.5ex}}%
	\stackon[1pt]{#1}{\tmpbox}%
}
\usepackage{longtable}

\def\beq{\begin{equation}}
	\def\eeq{\end{equation}}
\def\bea{\arraycolsep .1em \begin{eqnarray}}
	\def\eea{\end{eqnarray}}
\def\Tr{{\rm Tr}}
\def\tr{{\rm tr}}

\def\!!!{\stackrel{!}{=}}

\def\a{\alpha}
\def\b{\beta}

\def\eps{\varepsilon}

\def\STr{ {\rm STr } }

\def\M{ {\rm M} }

\def\nn{ \nonumber \\}

\def\eq#1{(\ref{#1})}

\def\s0#1#2{\mbox{\small{$ \frac{#1}{#2} $}}}
\def\0#1#2{\frac{#1}{#2}}

\def\grgl{\:\hbox to -0.2pt{\lower2.5pt\hbox{$\sim$}\hss}{\raise3pt\hbox{$>$}}\:}
\def\klgl{\:\hbox to -0.2pt{\lower2.5pt\hbox{$\sim$}\hss}{\raise3pt\hbox{$<$}}\:}

\begin{document}
	\title{Relational observables in Asymptotically safe gravity}
	\author{Alessio Baldazzi}\email{alessiobaldazzi93@libero.it}
	\address{\mbox{Scuola Internazionale di Studi Superiori Avanzati (SISSA) \& INFN, via Bonomea 265, 34136 Trieste, Italy}}
	\author{Kevin Falls}\email{kfalls@sissa.it}
	\address{\mbox{Scuola Internazionale di Studi Superiori Avanzati (SISSA) \& INFN, via Bonomea 265, 34136 Trieste, Italy}}
	\author{Renata Ferrero}\email{rferrero@uni-mainz.de}
	\address{\mbox{Institute of Physics (THEP), University of Mainz, Staudingerweg 7, D-55128 Mainz, Germany}}
	\vspace{1pt}
	\date{\today}
	\vspace{10cm}
	\begin{abstract}
		We introduce an approach to compute the renormalisation group flow  of relational observables in quantum gravity which evolve from their microscopic expressions towards the full quantum expectation value. This is achieved by using the composite operator formalism of the functional renormalisation group. These methods can be applied to a large class of relational observables within a derivative expansion for different physical coordinate systems.
		As a first application we consider four scalar fields coupled to gravity to represent the physical coordinate frame from which relational observables can be constructed. At leading order of the derivative expansion the observables are the inverse relational metric and the relational scalar curvature. We evaluate their scaling dimensions at the fixed point, both in the standard renormalisation group scheme and in the essential scheme. This represents the first steps to describe running observables within asymptotic safety; this treatment can be generalised to other observables constructed from  different tensors and in different physical coordinate systems.
	\end{abstract}
	\maketitle
	\tableofcontents
	\section{Introduction}\label{sec:intro}
	
	Asymptotic safety \cite{Weinberg} is by now a well established approach to quantum gravity which attempts to make sense of gravity as a fundamental quantum field theory \cite{perbook,rsbook}.
	A major critique of the current program is that it is not focused on observable quantities \cite{Donoghue:2019clr,Bonanno:2020bil,Knorr:2019atm}.
	However, within the functional renormalisation group (FRG) framework the flow equation for composite operators \cite{Pawlowski:2005xe} can be applied to investigate the scaling behaviour in quantum gravity\footnote{For an application to the Polyakov-loop in QCD see \cite{Herbst:2015ona}. } \cite{Pagani:2016dof,Becker:2018quq, Becker:2019fhi}. This formalism \cite{Pawlowski:2005xe,Pagani:2016pad,Pagani:2017tdr, Pagani:2020ejb} allows one to compute the expectation of observables by solving a flow equation where the flow parameter is the infra-red (IR) cut-off scale $k$. The initial condition at the ultra-violet (UV) cutoff scale $k = \Lambda$ is given by the expression of the observable as a function of the microscopic fields.  At the end point of the flow when $k=0$ the observable has evolved to the expectation value expressed as a function of the mean field. Crucially, in asymptotic safety the UV cutoff can be taken to infinity and hence the initial condition is taken defined in the limit $k \to \infty$ due to the existence of an interacting Reuter fixed point \cite{Reuter:1996cp, Reuter:2001ag,Souma:1999at, Falls:2014tra}. Thus, this formalism allows the computation of observables in an asymptotically safe theory of quantum gravity.
	
	Some applications have been performed considering  powers of the Ricci scalar integrated over spacetime \cite{Houthoff:2020zqy, Kurov:2020csd}.
	These terms would also naturally appear as terms in the effective action. On the other hand, the composite operator formalism allows one to compute the flow of terms which are not simply scalar quantities integrated over spacetime. Therefore it provides a formalism to compute quantities that are in principle measurable by a local experiment.
	
	For a quantum field theory in flat spacetime we could consider simply the values of the fields at a point in spacetime, or products thereof, as constituting observables. One of the greatest issues for a quantum theory of gravitation is how to generalise this notion of local observables \cite{Torre:1993fq, Torre:1994ef,Rovelli:1990ph, Rovelli:1990pi,Rovelli:2001bz,Dittrich:2004cb, Dittrich:2005kc,Dittrich:2015vfa,Westman:2007yx,Giddings:2005id, Donnelly:2016rvo,Hoehn:2020epv,Gielen:2018fqv}. Indeed, it is often stated that there are no local observables in a diffeomorphism invariant theory.
	However, this statement seems to contradict the fact that we are able to perform measurements of local fields: for example, measurements of the components of the Riemann tensor near the surface of the earth can be performed \cite{DeFelice:2010uvx}. It is then natural to assume such measurements are possible due to the presence of additional matter fields which provide a preferred frame of reference. Indeed, the matter fields  can be understood precisely as representing the readings on some physical coordinate scaffolding. More generally a `physical coordinate frame' constitutes four independent spacetime scalars which are  composed of the dynamical fields and their derivatives.
	In this way, one can construct a physical local reference frame and give physical meaning to the components of any tensor evaluated in this frame \cite{Komar:1958ymq,Bergmann:1960wb,Bergmann:1961wa}. In a cosmological setting, for example, a dense dust of particles \cite{Kuchar:1990vy,Brown:1994py}, which in the continuum limit becomes a continuous fluid, is commonly used to set up a physical coordinate system, and so give a physical meaning to continuous tensor field, which can be measured in some region of spacetime.
	Further examples of physical frames include curvature scalars formed from the Riemann tensor in pure gravity \cite{Komar:1958ymq,Bergmann:1960wb,Bergmann:1961wa,Khavkine:2015fwa}, massless scalar fields \cite{Brunetti:2013maa} and scalars derived from cosmological perturbation theory  \cite{Brunetti:2016hgw,Frob:2017gyj}. For a review and recent applications see \cite{Tambornino:2011vg, Ferrero:2020jts}.
	All measurements of tensor or scalar fields, referring to such physical coordinate systems, can be thought of as measurements of \textit{relational observables}.
	\bigskip
	
	Relational observables are therefore natural candidates for composite operators, which can be studied in asymptotic safety.
  Moreover at the Reuter fixed point the corresponding composite operators will have universal scaling exponents which, since they are observables, will not depend on unphysical elements of the scheme such as the gauge fixing conditions and choice of regulator.
		Physically these exponents should appear in the scaling behaviour of correlation functions of relational observables at small distances less than the Planck length where effects of the fixed point scaling are expected. 
		Once approximations are made however, we expect some dependence on the scheme. As such the computation of these exponents can serve as a way to compare with different approaches to quantum gravity and allow one to test the reliability of approximations. 
	The aim of this paper is to develop the appropriate formalism which allows these investigations. Furthermore, it is crucial that a consistent approximation scheme can be applied. In particular a general feature of the composite operator formalism is a mixing  of operators along the flow. For this purpose we develop a derivative expansion for relational observables which can be truncated in a consistent manner.
	

	
	\bigskip
	
	The paper is organised as follows. In Section \ref{sec:physcoordframe} we briefly review the ideas leading to the construction of relational observables through a physical coordinate system. In Section \ref{sec:compopefloweq} we review the functional integral formulation of composite operators. We derive the flow equation for the composite operators following the procedure of the FRG. Furthermore, we introduce a general formalism to translate the composite operator setting to relational observables and we comment about a natural criterion of choice of observables within a derivative expansion.  In Section \ref{sec:setup} we perform the first application to the flow of two relational observables corresponding to the inverse metric and the scalar curvature. In this example the physical coordinate system is composed of four massless scalar fields. The corresponding observables are those which are found at the leading order in the derivative expansion. In Section \ref{sec:as}, after having reviewed the matter flow within asymptotically safe gravity, we derive the flow of various physical matter systems (the Standard Model and modifications thereof) allowing for the spacetime metric to be renormalised. Here we consider two schemes, one in which the anomalous dimension of the metric is set to zero, and another one in which we adopt the minimal essential scheme, where we fixed the value of the cosmological constant by a renormalisation condition. We then compute the scaling of the chosen observables and calculate their scaling dimensions on the fixed points. In Section \ref{sec:general} we present the consistent extension of our application to the next order in the derivative expansion.
	
	In Appendices \ref{app:essential} and  \ref{app:scalartensor} we review the essential renormalisation group scheme \cite{Baldazzi:2021orb, Baldazzi:2021ydj} and we apply this to scalar tensor theories, deriving the flow equations. In Appendices \ref{app:hessian} and  \ref{app:calculations} we compute the Hessian of the observables and we perform the calculations leading to the numerical results for the scaling dimension of the couplings of the observables.
	\section{Physical coordinate frame and Relational observables}
	\label{sec:physcoordframe}
	\subsection{Physical coordinate frame}
	In order to construct relational observables we need a {\it physical coordinate frame} in which to evaluate tensors, or other diffeomorphism variant objects, such that when we transform both the tensor and the coordinate system the total transformation leaves the tensor invariant.
In other words a physical coordinate frame is a way to label a spacetime point $\mathcal{P}$ not by abstract coordinates $x^\mu$, but by the values which physical quantities take at $\mathcal{P}$. Thus, for example the scalar curvature $R(x)$ is not an observable since the coordinates $x$ do not mean anything physically. However, if the point $\mathcal{P}$ is specified as the point where a particular physical event happens, then the scalar curvature at $\mathcal{P}$ is an observable.
	If $\phi^a(x)$ denotes the full set of dynamical fields, which includes the components of the metric tensor $g_{\mu\nu}$ as well as matter fields, we construct a set of four scalars from the dynamical fields and their derivatives
	\beq \label{Xbar}
	\hat{X}^{\hat{\mu}}(x) = \hat{X}^{\hat{\mu}} ( \phi(x), \partial \phi(x),...)\,,
	\eeq
	which constitute our physical coordinate frame  with $\hat{\mu} = 0,1,2,3$. 
As a result a point $\mathcal{P}$ in spacetime is labelled by the values the scalar fields $\hat{X}^{\hat{\mu}}(x)$ take at $\mathcal{P}$.
	Since they are scalars, they transform, under a diffeomorphism of the dynamical fields $\phi \to \phi_\xi$, as $\hat{X}^{\hat{\mu}}(x)  \to  \hat{X}_\xi^{\hat{\mu}}(x) = \hat{X}^{\hat{\mu}}(\xi(x))$.
	Where here $\phi_\xi$ is the transformed dynamic field.
	In order that \eq{Xbar} constitutes a set of coordinates, they must constitute a diffeomorphism from the manifold $M$ to another manifold $\hat{M}$ with coordinates $\hat{x}^{\mu}$ which is the space of all values that the scalar fields $\hat{X}$ can take.
	Invertibility of the diffeomorphism  means that the equation
	\beq \label{x_bar}
	\hat{X}^{\hat{\mu}}(x) = \hat{x}^{\hat{\mu}}\,,
	\eeq
	can be solved for $x$. We denote the solution
	\beq
	x^{\mu} = X^{\mu}(\hat{x})\,,
	\eeq
	such that $X = \hat{X}^{-1}$ is the inverse of the map $\hat{X}$.
	Note that for this property to hold depends on the configuration $\phi$ and the choice of scalars $\hat{X}^{\hat{\mu}}$. Thus, a given choice of the scalars $\hat{X}^{\hat{\mu}}$ may define a physical coordinate system for a subset of all possible field configurations $\phi$.
	Given the set of scalars \eq{Xbar} one has a set of four covariant vectors
	\beq
	e^{\hat{\mu}}_{\mu}( x)
	=  \partial_{\mu}  \hat{X}^{\hat{\mu}}( x)\,.
	\eeq
	The frame fields $e_{\hat{\mu}}^{\mu}$ are the inverse of  $e^{\hat{\mu}}_{\mu}$ and transform as contravariant vectors on spacetime
	such that
	\beq
	e^{\hat{\mu}}_{\mu}(x) e_{\hat{\nu}}^{\mu}(x) = \delta^{\hat{\mu}}_{\hat{\nu}} \,,   \,\,\,\,\,\,\,\,\,\,\, e^{\mu}_{\hat{\mu}}(x) e^{\hat{\mu}}_{\nu}(x) = \delta^{\mu}_{\nu} \,.
	\eeq
	The frame fields $e_{\hat{\nu}}^{\mu}(x)$ can be obtained by differentiating the maps $X^\mu(\hat{x})$ and evaluating the derivative at $\hat{x} = \hat{X}(x)$. This follows since
	\beq
	\delta^\mu_\nu = \partial_\nu x^{\mu} =  e^{\hat{\mu}}_\nu(x)   \frac{\partial}{\partial \hat{x}^{\hat{\mu}}}X^{\mu}(\hat{X}(x))\,
	\eeq
	and hence we can write the frame field as
	\beq
	e^{\mu}_{\hat{\mu}}(x) = \frac{\partial}{\partial \hat{x}^{\hat{\mu}}} X^{\mu}(\hat{X}(x)) \,.
	\eeq
	Since we assume $e^{\hat{\mu}}_{\mu}$ is invertible it allows us to define an invariant volume element on space-time ${\rm d}^4x \tilde{e} $ where
	\beq
	\tilde{e} = \det e^{\hat{\mu}}_{\mu}\,.
	\eeq
	This allows us also to write the following identity between Dirac delta functions
	\beq
	\delta(X(\hat{x}),x)        = \tilde{e}(x) \, \delta (\hat{x},\hat{X}(x)   )\,.
	\eeq

	It is important to note that $  X^{\mu}(\hat{x})$ is itself a functional of the dynamical fields.
	Let us change the field configuration $\phi \to \phi' $ and we induce a change in  $\hat{X}(x) \to  \hat{X}'(x)$ obtained by replacing $\phi$ by $\phi'$ in  \eq{Xbar}. Since the coordinates $x$ are held fixed, we have that $X(\hat{x}) \to  X'(\hat{x})$  such that
	\beq \label{X'_to_X}
	X'( \hat{X}'(x)) =  X( \hat{X}(x)) =x\,.
	\eeq
	If we consider the case where $\phi'$  is obtained by a diffeomorphism of $\phi$ such that $\phi' = \phi_\xi$ , we have that $ \hat{X}'(x) = \hat{X}_\xi(x)=   \hat{X}(\xi(x))$.  We can then infer from \eq{X'_to_X}  that $X'(\hat{x})= X_\xi(\hat{x})$ is given by
	\beq \label{Transform_of_X}
	X_\xi(\hat{x})
	= (\xi^{-1}  \circ X)(\hat{x})\,.
	\eeq
	where $\xi^{-1}$ is the inverse function and $ f\circ g$ denotes the composition of two functions.
	
	\subsection{Relational observables}
	\label{subsec:relaobser}
	Given a field $\phi^a(x)$, let us denote its transformation under a diffeomorphism $\xi^\mu(x)$ by
	\beq
	\phi^a_\xi = T^a[\phi, \xi] \,,
	\eeq
	which is a functional of both the field $\phi$ and the diffeomorphism $\xi^{\mu}(x)$. For example a scalar field $\psi$ transforms to
	\beq
	\psi_\xi(x) = \psi(\xi(x)) \,,
	\eeq
	while components of the metric transform as
	\beq
	g_{(\xi) \mu\nu}(x) =  \partial_\mu \xi^{\lambda}(x)  \partial_\nu \xi^{\rho}(x)   g_{\lambda \rho}(\xi(x))\,.
	\eeq
	The functionals $T^a[\phi, \xi]$ satisfy the composition identity
	\beq \label{T_identity}
	T^a[\phi_\xi, \xi'] = T^a[\phi, \xi \circ \xi'] \,,
	\eeq
which follows from the properties that imply diffeomorphisms form a group where the group product is given by $(\xi' \cdot \xi)(x)  = \xi(\xi'(x))$.
	 For example if we consider a scalar $\psi(x)$, then applying first the diffeomorphism $\xi$ we have $  T[\psi, \xi] = \psi_\xi(x) = \psi(\xi(x))$, then applying  $\xi'$ we have $T[\psi_\xi, \xi'] = \psi_\xi(\xi'(x)) = \psi(\xi(\xi'(x)))$. Therefore the action of applying first $\xi$ and then $\xi'$ to $\psi$ is the same as 
	applying the composition in the reverse order $\xi\circ \xi'$.  
	
	To obtain the gauge invariant fields, we transform the dynamical fields $\phi^a(x)$ into the physical coordinate frame
	\beq \label{phihat}
	\hat{\phi}^{\hat{a}}(\hat{x}) = \phi^{\hat{a}}_{X} (\hat{x})
	\eeq
	such that it depends now on the coordinates $\hat{x}$.
	The fields $\hat{\phi}^{\hat{a}}(\hat{x})$ are then functionals $\hat{\phi}^{\hat{a}}[\phi]$ of the original fields and constitute a set of local observables at each point $\hat{x}$.
	As functionals of the fields $\phi$ we have that
	\beq
	\hat{\phi}^{\hat{a}} = T^{\hat{a}}[ \phi, X] \,.
	\eeq
	For example, the relational observable corresponding to the spacetime metric is given by the three equivalent expressions
	\bea
	\hat{g}_{\hat{\mu} \hat{\nu}}(\hat{x}) &=&  \partial_{\hat{\mu}} X^{\mu}(\hat{x})   \partial_{\hat{\nu}} X^{\nu}(\hat{x}) g_{\mu\nu}(X(\hat{x})) \\
	&=&    \int {\rm d}^4x \, \delta (x,X(\hat{x})   )   e^{\mu}_{\hat{\mu}}(x) e^{\nu}_{\hat{\mu}}(x)     g_{\mu \nu}(x)\\
	&=& \int {\rm d}^4x \,  \tilde{e}(x) \, \delta (\hat{x},\hat{X}(x)   )   e^{\mu}_{\hat{\mu}}(x) e^{\nu}_{\hat{\mu}}(x)     g_{\mu \nu}(x) \,.
	\eea
	To see that $\hat{\phi}^{\hat{a}} $ are observables, i.e., that they are diffeomorphism invariant,  we compute
	\bea
	\hat{\phi}^{\hat{a}}_\xi &=&  T^{\hat{a}}[ \phi_\xi, X_\xi]  \\
	&=&  T^{\hat{a}}[ \phi_\xi, \xi^{-1} \circ X  ]\\
	&=&   T^{\hat{a}}[ \phi, X  ] = \hat{\phi}^{\hat{a}} \,,
	\eea
	where to arrive at the second line we used \eq{Transform_of_X} and to  arrive on the third line we used  \eq{T_identity}.


	We can also take any composite operator $A[\phi]$ which transforms as $A[\phi_\xi] = A_\xi[\phi]$ and obtain a diffeomorphism invariant operator
	\beq
	\hat{A}[\phi] = A_X[\phi] = A[\hat{\phi} ]
	\eeq
	for any $A$ and any set of physical coordinates $\hat{A}[\phi] $ constitutes relational observable.
	For example, if $A= R(x)$ is the Ricci-scalar then
	\bea
	\hat{R}(\hat{x}) &=&    R(X(\hat{x}))  \nn
	&=&\int {\rm d}^dx \delta( X(\hat{x}), x)  R(x) \nn
	&=&\int {\rm d}^dx \, \tilde{e}(x) \, \delta (\hat{x},\hat{X}(x) )  R(x)
	\eea
	is the corresponding relational observable  which we dub the `relational Ricci-scalar'. While if $A = g^{\mu\nu}(x)$ is the inverse metric, then
	\beq
	\hat{g}^{\hat{\mu}\hat{\nu}}(\hat{x}) =   e^{\hat{\mu}}_\mu(X(\hat{x})) e^{\hat{\nu}}_\nu(X(\hat{x})) g^{\mu\nu}(X(\hat{x}))
	\eeq
	is the relational observable corresponding to the inverse metric or the `relational inverse metric', and
	\bea
	\hat{g}^{\hat{\mu}\hat{\nu}}(\hat{x})
	&=&\int {\rm d}^d x \,\delta(X(\hat x), x)\,g^{\mu\nu}(x) \nn
	&=&\int {\rm d}^d x \,\tilde e \, e^{\hat{\mu}}_\mu(X(\hat{x})) \,e^{\hat{\nu}}_\nu(X(\hat{x})) \,\delta(\hat x, \hat X(x))\,g^{\mu\nu}(x) \,.
	\eea
	Thus, given any diffeomorphism variant composite operator $A[\phi]$, then the relational observable $\hat{A}[\phi]$ can be dubbed the `relational $A$'. If the  variant composite operator $A^I(x)$ depends on the coordinates $x$ and some set of indices spacetime $I$ indices, e.g. $I =\mu \nu$ , the corresponding relational observable $\hat{A}^{\hat{I}}(\hat{x}) $ instead depend on $\hat{x}$ and hatted indices e.g. $\hat{I} = \hat{\mu}\hat{\nu}$.
	In general, if $A^I$ is a tensor, we can express the corresponding relational observable as
	\bea \label{Ahat}
	\hat{A}^{\hat{I}}(\hat{x}) = \int {\rm d}^4x \, \tilde{e}(x) \delta(\hat{x}, \hat{X}(x))     E^{\hat{I}}_I(x) A^I(x)\,,
	\eea
	where $E^{\hat{I}}_I(x)$ is a product of $e^{\mu}_{\hat{\mu}}$ and $e_{\mu}^{\hat{\mu}}$ depending on $I$, for example $E^{\hat{I}}_I(x) =e_{\mu}^{\hat{\mu}}e_{\nu}^{\hat{\nu}}$ when $A^I = g^{\mu\nu}$.
	
	Given a relational tensor $\widehat{A}^{\hat{I}}(\hat{x})$ we can also take derivative of the observable with respect to $\hat{x}$ to get a new relational observable.
	If $D_\mu$ denotes the covariant derivative compatible with $e^{\hat{\mu}}_\mu$
	\footnote{ {This means that $D_\mu$ is the covariant derivative compatible with the metric 
			$e^{\hat{\mu}}_\mu \delta_{\hat{\mu} \hat{\nu}} 
			e^{\hat{\nu}}_\nu$. Thus the connection for $\nabla_\mu$ is the Levi-Civita connection for $g_{\mu\nu}$ and the connection for $D_\mu$   is 
			the Levi-Civita connection for $e^{\hat{\mu}}_\mu \delta_{\hat{\mu} \hat{\nu}} 
			e^{\hat{\nu}}_\nu$.}} derivative with 
	and $D^n A^I_{\mu_1 ... \mu_n}  := D_{\mu_1} ... D_{\mu_n} A^I$ is the $n$th derivative of the gauge variant  composite operator $A[\phi]$, then one can show that
	\beq
	\partial_{\hat{\mu}_1} ...\partial_{\hat{\mu}_n} \hat{A}^{\hat{I}} =  \widehat{D^n A}_{\hat{\mu}_1...\hat{\mu}_n}^{\hat{I}}
	\eeq
	simply by differentiating \eq{Ahat}, using that $\partial_{\hat{\mu}} \delta(\hat{x}, \hat{X}(x)) = -  e^{\mu}_{\hat{\mu}} D_\mu \delta(\hat{x}, \hat{X}(x) ) $ and integrating by parts repeatedly.
	
	Up till now we have assumed that $A$ is a tensor. However, we can consider relational observables which are not related to tensors. For example  the Christoffel symbol $\Gamma^\lambda_{\mu\nu}$ transforms inhomogeneously. As a  result the relational Christoffel symbol is given by
	\bea
	\Gamma_{\hat{\mu}\hat{\nu}}^{\hat{\rho}}(\hat{x}) &=&  \int {\rm d}^4x \, \tilde{e} \, \delta(\hat{x}, \hat{X}) \, e^{\mu}_{\hat{\mu}} e^{\nu}_{\hat{\nu}}  \left( e_{\rho}^{\hat{\rho}}  \Gamma_{\mu \nu}^{\rho}   -    \partial_{\mu} \partial_\nu \hat{X}^{\hat{\rho}}  \right) \nn
	&=& - \int {\rm d}^4x \, \tilde{e}  \, \delta(\hat{x}, \hat{X})  e^{\mu}_{\hat{\mu}} e^{\nu}_{\hat{\nu}}    \nabla_{\mu} \nabla_\nu \hat{X}^{\hat{\rho}}  \,,
	\eea
	which is diffeomorphism invariant. However, we observe that this is the relational observable corresponding to the tensor $-\nabla_{\mu} \nabla_\nu \hat{X}^{\hat{\rho}}$.

	From these considerations we see that there are many different relational observables that one can construct since there is a immense freedom in both choosing the physical coordinate system and choosing which composite operator $A[\phi]$ to transform into the chosen coordinate system.
	From this point of view, there is no lack of observables in quantum gravity, but rather perhaps a lack of determining which observables are most relevant amongst the large number of relational observables we can consider in principle.
	
	
	\section{Composite operator flow equation}
	\label{sec:compopefloweq}
	In many applications of  quantum field theory (QFT) we are interested in correlation functions of the fundamental fields $\hat\chi$. However, sometimes the physical quantities of interest may be composite operators $\hat{\mathcal{O}}(x)$ which are functions of the field and its derivatives. For example, we could consider the field to some power $n$ i.e. $\hat{\mathcal{O}}(x) = \hat\chi^n(x)$ .
	In the functional integral formulation of standard QFT one can treat composite operators $\hat{\mathcal{O}}(x)$ coupling them to external sources by adding to the microscopic  action a term
	\beq
	\int {\rm d}^dx \, \hat{\mathcal{O}}(x)  \eps (x) \,,
	\eeq
	where $\eps(x)$ denotes the source at the point $x$.
	In this way one obtains insertions of composite operators in correlation functions by taking suitable functional derivatives of the path integral with respect to the source \cite{Pagani:2016pad,Pagani:2016dof}. For example the expectation value of the operator ${\mathcal{O}}(x)$ is given by
	\bea
	\langle \hat{\mathcal{O}}(x)\rangle \; = -\frac{\delta}{\delta \eps (x)} \int  ({\rm{d}} \hat{\chi}) ~\,{\rm e}^{-S[\hat{\chi}] - \eps \cdot \hat{\mathcal{O}}[\hat{\chi}]}\Bigg|_{\eps = 0} \,.
	\eea
	Then, we define the generating functional $W [J, \eps]$ for the connected Green's functions of the fields and the connected correlation functions of the composite operators by
	\bea
	{\rm e}^{W[J, \eps]} \; \equiv \int ({\rm{d}} \hat{\chi}) ~\, {\rm e}^{-S[\hat{\chi}] - \eps \cdot \hat{\mathcal{O}}[\hat{\chi}]+ J\cdot \hat\chi} \,.
	\eea
	The associated effective action is obtained via a Legendre transform with respect to $J$ to obtain a functional $\Gamma[\phi, \eps]$ where $\phi$ is the expectation value of the field $\hat \chi$ in the presence of the source $J$. The expectation of the composite operator is then given by expanding the effective action around $\eps = 0$
	\beq
	\Gamma[\phi, \eps]   = \Gamma[\phi] + \int   {\rm d}^dx \, \eps(x)  \langle \hat{\mathcal{O}}(x)\rangle_\phi   +O(\eps^2) \,,
	\eeq
	which depends on the value of the the mean field.

	The effective action  $\Gamma[\phi, \eps]$ can  also be modified  to allow for a regulator $\mathcal{R}_k$ which suppresses infra-red fluctuations of the fields $\hat\chi$ around their mean values.
	In particular, the effective average action  $\Gamma_k [\phi, \eps]$ can be defined by its functional integro-differential  representation
	\beq
	e^{-\Gamma_k [\phi, \eps]} = \int ({\rm{d}} \hat{\chi}) ~ {\rm e}^{-S[\hat{\chi}] - \eps \cdot \hat{\mathcal{O}}[\hat{\chi}] + (\hat \chi - \phi) \cdot \frac{\delta}{\delta \phi}  \Gamma_k [\phi, \eps]  - \frac{1}{2}(\hat \chi - \phi) \cdot \mathcal{R}_k  (\hat\chi - \phi) } \,.
	\eeq
	Note that $\mathcal{R}_k$ is an additive IR cut-off which suppresses fluctuations below momentum scales $p^2\simeq k^2$ and vanishes for momenta $p^2 \gg k^2$.
	The $k$ and $\phi$ dependent expectation value is given by  $\mathcal{O}_k(x) \equiv \langle \hat{\mathcal{O}}(x) \rangle_{\phi,k} $ is again given by expanding  $\Gamma_k [\phi, \eps]$  to first order in $\eps$
	\beq
	\Gamma_k[\phi, \eps]   = \Gamma_k[\phi] + \int   {\rm d}^dx \, \eps(x) \mathcal{O}_k(x) + O(\eps^2) \,.
	\eeq
	The exact FRG equation for $ \Gamma_k[\phi, \eps] $ is given by the Wetterich-Morris equation \cite{Wetterich:1992yh, Morris:1993qb}
	\beq
	\partial_t  \Gamma_k[\phi, \eps]  =  \frac{1}{2}\Tr \left[\left(\Gamma_k^{(2,0)}[\phi,\eps]  + \mathcal{R}_k\right)^{-1}\partial_t \mathcal{R}_k\right]\,,
	\eeq
	where $t = \log(k/k_0)$ is the RG time.
	Now comparing order by order in $\eps$, the flow equation for composite operators is given by \cite{Pawlowski:2005xe}
	\bea
	\int   {\rm d}^dx  \, \eps \partial_t  \mathcal{O}_k = -\frac{1}{2} \Tr\left[\left(\Gamma_k^{(2)} + \mathcal{R}_k\right)^{-1}\left(  \int   {\rm d}^dx \,  \eps \mathcal{O}_k^{(2)}\right)\left(\Gamma_k^{(2)} + \mathcal{R}_k\right)^{-1}\partial_t \mathcal{R}_k\right] \,,
	\label{flow_cobis}
	\eea
	where $\mathcal{O}^{(2)}_k$ is the Hessian of the composite operator. The flow equation for the composite operator  $ \mathcal{O}_k$ must be supplied by an initial condition at $k = \Lambda$. Since, in the limit $k \to \infty$ we have that
	\beq
	\lim_{k \to \infty} \mathcal{O}_k  =   \hat{\mathcal{O}}|_{\hat{\chi} \to \phi}
	\eeq
	setting the initial condition for $\Lambda \to \infty$ specifies which observable $\hat{\mathcal{O}} $ we are taking the expectation value of.
	Then by following the flow to $k=0$ we obtain the expectation value
	\beq
	\mathcal{O}_{0} = \langle  \hat{\mathcal{O}} \rangle\,.
	\eeq
	In this manner solving the flow equation for the composite operator gives us a concrete method to compute expectation values of observables.

	To concretely solve equation, some approximation must be implemented. In particular, one may expand the composite operator $\mathcal{O}_k $ in a basis of $k$-independent operators
	\beq \label{Ok_expansion}
	\mathcal{O}_k(x) = \sum_i  a_i(k)  \mathcal{O}_i(x)\,.
	\eeq
	One can show that the scaling operators of the theory have dimensions, quantum corrections included, given by the eigenvalues of the matrix
	\bea
	S_{ij}=d_i \delta_{ij} + \gamma_{ij}  \,, \,\,\,\,\,\,   \partial_t a_j = \sum_i a_i \gamma_{i j} \,,
	\eea
	where $S_{ij}$ is a function of the couplings included in the effective action $\Gamma_k$.
	The crucial matrix $\gamma_{ij}$ can be directly found, inserting the expansion \eqref{Ok_expansion} in the flow equation and expanding the trace
	\bea \label{CO_trace}
	-\frac{1}{2} \Tr\left[\left(\Gamma_k^{(2)} + \mathcal{R}_k\right)^{-1}\left( {\rm d}^dx  \, \eps(x) \mathcal{O}_i^{(2)}(x) \right)\left(\Gamma_k^{(2)} + \mathcal{R}_k\right)^{-1}\partial_t \mathcal{R}_k\right]
	\label{CO} = \sum_j \gamma_{ij} \int {\rm d}^dx \,  \eps(x)  \mathcal{O}_j(x) \,.
	\eea
Note that in the rhs of \eqref{CO_trace} no terms with derivatives of  $\eps(x)$ appear since we can always integrate by parts such that all derivatives are included in $\mathcal{O}_j(x)$.
	
	\subsection{Role of the source}
	\label{sec:rolesource}
	Before applying the composite operator formalism to relational observables, let us consider the role of the source $\eps(x)$ in calculating $\gamma_{ij}$. In particular, suppose we choose to make the source a constant $\eps(x) = \eps$. Then, we can factor our the source and it appears as a source for the observable integrated over spacetime
	\beq
	\int {\rm d}^dx \, \eps \mathcal{O}_k(x) = \eps  \sum_i  a_i(k)   \int {\rm d}^dx \,  \mathcal{O}_i(x)
	\eeq
	such that if $\int {\rm d}^dx   \mathcal{O}_i(x)$ are linearly dependent then we are not able to determine all components of $\gamma_{ij}$. Essentially this means that, in the absence of boundaries, the flow of all operators which are total derivatives are lost by taking the source to be constant. Let us therefore consider a basis where all linearly independent {\it boundary operators} of the form $\mathcal{O}_i = \partial_\mu \mathcal{O}^\mu_i$ is a subset of the basis which is completed by taking a set of { \it bulk operators} which are linearly independent of boundary operators. Then, if $i$ is a boundary  index and $j$ is a bulk index, then $\gamma_{ij}=0$. To see this, note that
	\beq
	\int {\rm d}^dx \, \eps(x) (\partial_\mu \mathcal{O}^\mu_i)^{(2)} = - \int {\rm d}^dx  (\partial_\mu \eps(x))  \mathcal{O}^\mu_i\,^{(2)} \,,
	\eeq
	and thus if $\mathcal{O}_i(x) =  \partial_\mu \mathcal{O}^\mu_i$ is inserted into the lhs \eq{CO_trace}, then rhs must be given by  $ - \sum_{j={\rm boundary \, indices}} \gamma_{ij} \int {\rm d}^dx  \partial_\mu \eps(x)  \mathcal{O}_j^\mu(x)$. A consequence of this is that by taking the source to be constant, we are still able to compute the flow of the bulk terms in a consistent manner. Allowing the source to be non-constant, we are then able to compute the flow of the boundary operators in addition.


	\subsection{Application to relational observables}
	To apply this formalism to a relational observables $\hat{A}^{\hat{I}}(\hat{x})$ and their derivatives, we can consider a composite operator $\mathcal{O}_k(\hat{x})$ to be any function of relational observables at the point $\hat{x}$. This can be decomposed as
	\beq
	\mathcal{O}_k(\hat{x}) =  \sum_i  a_{\hat{I}_i}(k) \hat{A}^{\hat{I}_i}_i(\hat{x})\,,
	\eeq
	where the index $i$ runs over different relational observables, taking into account that the observables can mix under the RG flow.
	We note that in general each component of each observable can have a separate coupling $a_{\hat{I}_i}(k)$.
	The source term can then be written as an integral over $\hat{x}$ as \eq{Ahat}
	\beq
	\int {\rm d}^4\hat{x} \,  \eps(\hat{x}) \mathcal{O}_k(\hat{x})  =  \sum_i  a_{\hat{I}_i}(k)  \int {\rm d}^4\hat{x}  \, \eps(\hat{x})   \hat{A}^{\hat{I}_i}_i(\hat{x})	\,,
	\eeq
	however we can also express the source term as an integral over the coordinates $x$ by using
	\beq \label{Ok_x_integral}
	\int {\rm d}^4\hat{x}  \, \eps(\hat{x}) \mathcal{O}_k(\hat{x})  = \int {\rm d}^4x \, \tilde{e}(x)  \, \eps(\hat{X}(x))  \sum_i  a_{\hat{I}_i}(k)   E^{\hat{I}_i}_{i I_i}(x)   A^{I_i}_i(x) \,.
	\eeq
	Now since $ \eps(\hat{x})$ is independent of the dynamical fields, it is clear that variations of the source term with respect to dynamical fields must be proportional to the undifferentiated source $\eps(\hat{x})$. However, in the form \eq{Ok_x_integral} $ \eps(\hat{X}(x))$ depends on the dynamical fields through the physical coordinate system. Thus, in computing the Hessian of the observable with a non-constant source we pick up terms from the variation of $ \eps(\hat{X}(x)) $ and thus are proportional to the derivative of the source. This is consistent since by using the identity
	\beq
	\frac{\partial}{\partial \hat{X}^{\hat{\mu}}(x)}  \eps(\hat{X}(x))  =  e^{\mu}_{\hat{\mu}}(x)  \partial_\mu \eps(\hat{X}(x)) \,,
	\eeq
	and integrating by parts we can always write the variation such that it is proportional to $\eps(\hat{X}(x))$.
	
	According to our considerations in the last section, we can consider constant sources if we concentrate on relational observables which are linearly independent of total derivatives $\hat{A}^{\hat{I}_i}_i(\hat{x}) =  \partial_{\hat{\mu}} \hat{A}^{\hat{I}_i \hat{\mu}}_i(\hat{x})$. Indeed, working with integrals over $\hat{x}$ the arguments of the last section go through unchanged. Let's note that when we switch to the form \eq{Ok_x_integral} we have that
	\beq
	\int {\rm d}^4\hat{x}  \, \eps(\hat{x})  \partial_{\hat{\mu}} A^{\hat{I}}(\hat{x}) =  \int {\rm d}^4x 	\, \tilde{e}(x)  \, \eps(\hat{X}(x))  E^{\hat{I}}_{ I}(x)  e^{\mu}_{\hat{\mu}} D_\mu A^{I}(x)\,,
	\eeq
	where the lhs is a boundary term when $ \eps(\hat{X}(x)) $ is a constant. So as with any composite operator the computation of scaling dimensions of the subset of observables which are not total derivatives can be calculated at constant source.
	
	Taking the source constant, we can define the {\it relational EAA} by
	\beq \label{relational_action}
	\boxed{	\Gamma_k^{\rm rel.} \equiv   \int {\rm d}^4x \, \tilde{e}(x)    \mathcal{L}^{\rm rel. }_k(x)     :=      \int {\rm d}^4x \, \tilde{e}(x)   \sum_i  a_{\hat{I}_i}(k)   E^{\hat{I}_i}_{i I_i}(x)   A^{I_i}_i(x)  } \,,
	\eeq
	where we note that the defining feature that separates terms in $\Gamma_k^{\rm rel.} $ from terms in the standard effective action is that the former is written as an integral over  $x$ with the volume element given by $\tilde{e}(x)$ rather than the usual density $\sqrt{\det g}$. Thus, for constant source, we can take the total EAA to be given by
	\bea
	\Gamma_k[g,...; \eps] = \Gamma_k[g,...] +  \eps  \,  \Gamma_k^{\rm rel.}[g,...]  + O(\eps^2) \,.
	\eea
	Implicit in the construction of the relational EAA is that $\tilde{e}$ is non-singular. The form \eq{relational_action}  gives us two different starting points. On the one hand, we can consider a basis of relational observables and give each one a source. On the other hand, we can simply write down some form for $\mathcal{L}^{\rm rel. }_k(x)$ as a basis of scalars and the non singular nature of  $\tilde{e}$ means we can construct these scalars with the vectors $e^{\mu}_{\hat{\mu}}$ and there derivatives.
	
	For a non-constant source we instead have the more general form
	\bea
	\Gamma_k[g,...; \eps] = \Gamma_k[g,...] +  \int {\rm d}^4x 	\, \tilde{e}(x)  \, \eps(\hat{X}(x))  \mathcal{L}^{\rm rel. }_k(x) + O(\eps^2) \,,
	\eea
	which allows for the calculation of scaling dimensions for relational observables which are total derivatives and hence do not appear in the relational EAA.

	\subsection{Derivative expansion for the relational EAA}
	
	Applying the composite operator formalism to the relational  EAA (to order $\eps$) we have that
	\bea
	\boxed{\partial_t \Gamma_k^{\rm rel.}  = -\frac{1}{2} \Tr\left[\left(\Gamma_k^{(2)} + \mathcal{R}_k\right)^{-1}\left(\Gamma_k^{{\rm rel.}(2)} \right)\left(\Gamma_k^{(2)} + \mathcal{R}_k\right)^{-1}\partial_t \mathcal{R}_k\right]} \,
	\label{flow_co}
	\eea
	where $\Gamma_k^{{\rm rel.}(2)}$ is the Hessian of the relational EAA. An important question is how to close approximations to this equation in a consistent manner. A natural non-perturbative choice is to take  $\mathcal{L}^{\rm rel. }_k(x)$ to have a derivative expansion and be approximated by terms with only a finite number of derivatives, however this only makes sense if $\nabla_\mu \hat{X}^{\hat{\mu}}$ is polynomial in derivatives since otherwise $\tilde{e}$ would be non-polynomial. This provided, we pick  $\hat{X}^{\hat{\mu}}$ only involve a finite number of derivatives, we can close our approximation for the relational EAA by also picking terms with a up to a finite number of derivatives in $\mathcal{L}^{\rm rel. }_k(x)$. This then gives us a natural basis of relational observables which we project onto when we use the derivative expansion. In particular, these observables are formed from tensors $A^I$ with all indices in the upper position, such as $g^{\mu\nu}$, $R$,  $R^{\mu\nu}$ etc., since the corresponding $E^{\hat{I}}_I$ are products of $\nabla_\mu \hat{X}^{\mu}$ and thus finite order in derivatives.  Each order $s$ of the derivative expansion then corresponds to keeping only terms with up to $s$ derivatives in  $\mathcal{L}^{\rm rel. }_k(x)$.   At each finite order in the derivative expansion, we do not find relational observables corresponding to tensors with lower indices, e.g. $g_{\mu\nu}$, which are non-polynomial in derivatives.

	\section{First application}
	\label{sec:setup}
	We are now ready to apply the composite operator formalism to the investigation of relational observables in asymptotic safety.
	Here we employ the background field approximation where
	\beq
	\Gamma_k = \bar{\Gamma}_k + \Gamma_{{\rm gh} k} + \Gamma_{{\rm gf} k}
	\eeq
	is split into a diffeomorphism invariant part $\bar{\Gamma}_k$, a gauge fixing part $\Gamma_{{\rm gf} k}$  and a ghost action   $\Gamma_{{\rm gh} k}$. Rather than working with pure gravity we take $\bar{\Gamma}_k$  to depend on matter fields from which we construct the physical coordinate system $\hat{X}^{\hat{\mu}}(x)$ that enters the relational observables. A simple example is the inclusion of a set of four massless scalar fields minimally coupled to gravity (see \cite{Percacci:2003jz} for gravity coupled to matter in asymptotic safety and \cite{Laporte:2021kyp} for a more recent analysis of the fixed point existence):\footnote{Here the anomalous dimension of the matter fields has been neglected and only one-loop results have been considered. In a more recent paper \cite{Dona:2013qba} the running of the anomalous dimension of all the involved matter fields has been taken into account.}
	\beq
	\bar{\Gamma}_k = \int_x \sqrt{{\rm det}g} \left(  -\frac{1}{16  \pi G_N}  (R-2\Lambda) + \frac{1}{2}  \delta_{\rm{AB}} g^{\mu\nu} \partial_{\mu}\varphi^{\rm{A}} \partial_{\nu}	 \varphi^{\rm{B}}\right) + \text{Additional Matter} \,,
	\label{action}
	\eeq
	where the index $\rm{A},\rm{B}, ... = 1,2,3,4$ run over the internal space of the scalar fields and $ \delta_{\rm{AB}} $ is a flat field space metric. For the theory described by \eqref{action} there are in principle many choices for the scalars $\hat{X}^{\hat{\mu}}(x)$,  here we identify the physical coordinate system with the four scalars $\varphi^{A}(x)$ such that
	\beq \label{Xhat_eq_varphi}
	\hat{X}^{\hat{\mu}}(x) = \varphi^{\hat{\mu}}(x) \,.
	\eeq
	In this case the physical coordinate system is just composed for fundamental scalars rather than composite fields.
	After having made the identification \eqref{Xhat_eq_varphi}, we simply denote the scalar fields by $ \hat{X}^{\hat{\mu}}(x)$ and always use the hatted indices $\hat{\mu}$ such that the field space metric is $\delta_{\hat \mu \hat \nu}$. Let us note that the action \eqref{action} is invariant under shifts $ \varphi^A(x)  \to \varphi^A(x) + c $ and under a global $O(4)$ symmetry. For simplicity, we consider observables that are invariant under these symmetries also. Since these symmetries are not broken by the regularisation, the flow of symmetric composite operators is closed. One can consider observables which break these global symmetries however we don't do so here.
	
	For the gauge fixing action we take
	\beq
	\Gamma_{{\rm gf} k}[g;\bar{g}] = \frac{1}{2 } \int {\rm d}^d x \sqrt{\det \bar{g}} F^{\nu}  \bar{g}_{\mu\nu} F^{\mu} \,,
	\eeq
	where  we use the background covariant harmonic gauge
	\bea
	F^\mu
	&=& \frac{1}{\sqrt{16 \pi G_k}}  \left(  \bar{g}^{\mu \lambda} \bar{g}^{ \nu \rho}  - \frac{1}{2} \bar{g}^{\nu\mu}   \bar{g}^{\rho \lambda}     \right)  \bar{\nabla}_\nu  g_{\lambda \rho}\, .
	\eea
	This leads to the ghosts operator
	\beq
	\mathcal{Q}^{\mu}\,_{\nu} c^{\nu} \equiv \mathcal{L}_{c} F^{\mu} =   \frac{1}{\sqrt{16 \pi G_k}} \left(  \bar{g}^{\mu \lambda} \bar{g}^{ \nu \rho}  - \frac{1}{2} \bar{g}^{\nu\mu}   \bar{g}^{\rho \lambda}     \right)  \bar{\nabla}_\nu ( g_{ \rho \sigma} \nabla_{\lambda} c^{\sigma}  + g_{ \lambda \sigma} \nabla_{\rho} c^{\sigma}) \,,
	\eeq
	which enters the ghost action
	\beq
	\hat{\Gamma}_k[g, c ,\bar{c};\bar{g}] = \int {\rm d}^d x \sqrt{\det \bar{g}}  \bar{c}_\mu \mathcal{Q}^{\mu}\,_{\nu} c^{\nu}  \,.
	\eeq


	\subsection{Inverse relational metric and relational curvature}
	In this work we use the composite operator formalism to compute the scaling dimensions of relational observables corresponding to the inverse metric and the scalar curvature where the physical coordinates $\hat{X}^{\mu}(x)$ are taken to be a set massless minimally coupled scalars. To do so, it suffices to consider the integral of the relational observables over field space. In particular, we take the relational EAA to be linear in the relational inverse metric and the relational curvature integrated over the $\hat{x}$ coordinate
	\beq
	\boxed{\Gamma_k^{\rm rel.} =   \int {\rm d}^4 \hat{x}  \left( \alpha_0(k)     + \alpha_R(k)  \hat{R}(\hat{x}) + \alpha_1(k) \delta_{\hat{\mu}\hat{\nu}}    \hat{g}^{\hat{\mu}\hat{\nu}}(\hat{x}) \right)} \,,
	\label{ourgammarel}
	\eeq
	where $\delta_{\hat{\mu}\hat{\nu}} $ is the Kronecker delta which is the same the metric on field space that appears in the EAA.
	In this case, the sources which couple to the relational inverse metric and the relational curvature are taken to be equal to $\delta_{\hat{\mu}\hat{\nu}} $ and $1$ respectively.
	We note that the first term represents the volume of the field space which  is independent of the dynamical fields. However, we see that $ \alpha_0(k) $ has a flow which depends on the other two couplings $ \alpha_1(k)$ and $ \alpha_R(k)$ which couple to the relational observables. We can then rewrite $\Gamma_k^{\rm rel.}$ in the form \eq{relational_action} where by we have
	\beq
	\Gamma_k^{\rm rel.} =   \int {\rm d}^4 x \, \tilde{e}  \left( \alpha_0(k)  + \alpha_R(k) R  + \alpha_1(k) \delta_{\hat{\mu} \hat{\nu}} g^{\mu\nu} (\partial_{\mu} \hat{X}^{\hat{\mu}}) (\partial_{\nu} \hat{X}^{\hat{\nu}})   \right)\,.
	\label{relactionorder2}
	\eeq
	We note that the corresponding $\mathcal{L}^{\rm rel. }_k(x)$ contains terms with up to two derivatives of the field. Apart from the difference in the volume element the terms in  $\Gamma_k^{\rm rel.}$ are the same as $\Gamma_k$ indicating a consistency in the approximation of both functionals. Moreover, $\mathcal{L}^{\rm rel. }_k(x)$ contains all terms with up to two derivatives compatible with the symmetries of the action $\Gamma_k$. Thus, here we are working at order $\partial^2$ in the derivative expansion.
	
	Let us remark that since the relational action includes the volume element $\tilde{e}$ one has that
	\beq
	\int {\rm d}^4 x \, \tilde{e} \delta_{\hat{\mu} \hat{\nu}} g^{\mu\nu} (\partial_{\mu} \hat{X}^{\hat{\mu}}) (\partial_{\nu} \hat{X}^{\hat{\nu}})  \neq   - \int {\rm d}^4 x \, \tilde{e} \delta_{\hat{\mu} \hat{\nu}} \hat{X}^{\hat{\mu}}  \Box \hat{X}^{\hat{\nu}}	\,,
	\eeq
	since integrating by parts leads to terms where the covariant derivatives act on $ \tilde{e} $.
	However, the term on the rhs, which can be expressed as
	\beq
	- \int {\rm d}^4 x \, \tilde{e} \delta_{\hat{\mu} \hat{\nu}} \hat{X}^{\hat{\mu}}  \Box \hat{X}^{\hat{\nu}}	 = 	 \int {\rm d}^4 \hat{x} \, \delta_{\hat{\mu} \hat{\nu}} \hat{x}^{\hat{\mu}}  \hat{g}^{\hat{\rho} \hat{\lambda}} \hat{\Gamma}^{\hat{\nu}}_{\hat{\rho} \hat{\lambda}}\,,
	\eeq
	is not symmetric under the shift symmetry of the scalar field and so we do not consider it here.

	\subsection{Flow of the relational observables}
	\label{sec:results}

	We are now ready to compute the flow of the relational action \eqref{relactionorder2} by using \eqref{flow_co} where the EAA is given by \eqref{action}. We use a type I cutoff for which the regulated Hessian is
	\beq
	\Gamma^{(2)}_k + \mathcal{R}_k = \Gamma^{(2)}_k(-\nabla^2 \to P_k) \,,
	\eeq
	where $P_k= -\nabla^2 + \mathcal{R}_k(-\nabla^2)$.
	
	The trace in \eqref{flow_co} is split into two contributions: the trace over the graviton fluctuations and the trace over the scalar fluctuations. Within our approximation, there are no contributions from the ghosts or any additional matter fields. An improved approximation would include terms in \eqref{relactionorder2} containing ghosts and additional matter fields since such terms would be generated by the flow equation.
	
	All the intermediate steps which lead to the following expressions can be found in Appendix \ref{app:calculations}. Here we use the off-diagonal heat kernel techniques to evaluate the traces. We also allow for a non-zero anomalous dimension of the metric tensor $g_{\mu \nu}$ and the scalar fields $X^\alpha$. This has two effects on the flow Equations \eqref{flow_co}. In the lhs we replace
	\beq
	\partial_t \Gamma_k^{\textrm{rel.}} \to \partial_t \Gamma_k^{\textrm{rel.}} -\frac{1}{2}\hat \eta_k \int \textrm{d}^4x \hat X^{\hat \mu} (x) \frac{\delta \Gamma_k^{\textrm{rel.}}}{\delta  \hat X^{\hat \mu} (x) }+\gamma_g \int \textrm{d}^4x  g_{\mu \nu}(x) \frac{\delta \Gamma_k^{\textrm{rel.}}}{\delta  g_{\mu \nu}}(x) \,,
	\eeq
where $\hat{\eta}$ is the anomalous dimension of scalar fields and  $\gamma_g$ accounts for the anomalous dimension of the metric tensor. 
	In the rhs the derivative of the cutoff for the metric fluctuations is replaced by
	\beq
	\partial_t \mathcal{R}_k \to \partial_t \mathcal{R}_k  + 2 \gamma_g \mathcal{R}_k\; ,
	\label{R_kg}
	\eeq
	while the derivative of the cutoff for the scalar fluctuations is replaced by
	\beq
	\partial_t \mathcal{R}_k \to \partial_t \mathcal{R}_k  -  \hat \eta_k \mathcal{R}_k \,.
	\eeq

	The flow of the coefficients $\a_0$, $\a_R$ and $\a_1$ is given by
	\begin{align}
		&\partial_t\alpha_0 = -\frac{\alpha_1}{2 \pi^2}
		Q_3\left[ \frac{\left(\partial_t -\hat\eta_k \right) \mathcal{R}_k }{ P_k^2} \right]
		-\frac{ 12\alpha_R\,G_N}{\pi}
		Q_3\left[ \frac{\left(\partial_t -\eta_\phi \right) \mathcal{R}_k }{\left( P_k -2\Lambda \right)^2} \right]\,,
		\\
		&\partial_t\alpha_R  -\gamma_g \, \alpha_R=
		-\frac{\alpha_1}{24 \pi^2}  \,
		Q_2\left[ \frac{\left(\partial_t -\hat\eta_k \right) \mathcal{R}_k }{ P_k^2} \right]
		-\frac{5\alpha_R\,G_N}{3\pi} \,
		Q_2\left[ \frac{\left(\partial_t -\eta_\phi \right) \mathcal{R}_k }{\left( P_k -2\Lambda \right)^2} \right]
		+ \frac{12\alpha_R\, G_N}{\pi}
		Q_3\left[ \frac{\left(\partial_t -\eta_\phi \right) \mathcal{R}_k }{ (P_k-2\Lambda)^3} \right]\,,	\label{floweq1}	\\
		&\partial_t\alpha_1-\left( \gamma_g + \hat\eta_k \right) \alpha_1=
		-\frac{4\alpha_1\,G_N}{\pi} \,
		Q_2\left[ \frac{\left(\partial_t -\eta_\phi \right) \mathcal{R}_k }{\left( P_k -2\Lambda \right)^2} \right]
		- \frac{5\alpha_1\, G_N}{8\pi}
		Q_3\left[ \frac{\left(\partial_t -\hat\eta_k \right) \mathcal{R}_k }{ P_k^3} \right]
		-6 \alpha_R\, G_N^2
		Q_3\left[ \frac{\left(\partial_t -\eta_\phi \right) \mathcal{R}_k }{ (P_k-2\Lambda)^3} \right]\,.
	\end{align}
	where for a function of a Laplacian $W(-\nabla^2)$ we defined the $Q$-functional
	\beq
	Q_n[W]= \frac{1}{\Gamma(n)} \int_0^{\infty} {\rm d}z\; z^{n-1}W(z)\;,
	\eeq
	and $\eta_\phi$ is the anomalous dimension of the graviton fluctuations
	\beq
	\eta_{\phi} = \eta_N -2\gamma_g\; \quad \text{where} \quad \eta_N = \frac{\partial_tG_k}{G_k}.
	\eeq
	The first term originates from the cutoff dependence on $G_k^{-1}$ and the second term arises due to the replacement \eqref{R_kg}.
The anomalous dimension of the graviton fluctuations $\eta_\phi$ can be understood as the anomalous dimension of the graviton field $\phi_{\mu \nu}$ defined by 
		\beq
		g_{\mu\nu} = \delta_{\mu\nu} + \sqrt{32 \pi G_k} \phi_{\mu\nu}
		\eeq
		where $\delta_{\mu\nu}$ is the flat metric and the factor $ \sqrt{32 \pi G_k}$ ensures that $\phi_{\mu\nu}$ has a canonically normalised kinetic term.
	
	\bigskip
	
	\underline{Remark}: In four dimensions: the scalars have the canonical dimension $[\hat X] = 1$, the metric has dimension $[g_{\mu \nu} = -2]$, s.t. the determinant $[\tilde e] = 4$ and the inverse relational metric $[\hat \delta_{\hat \mu \hat \nu}g^{\mu \nu}\partial_\mu \hat X^{\hat \mu} \partial_\nu \hat X^{\hat \nu}] = 4$. In 4 dimensions then the coefficients $[\a_0] = -4$, $[\a_R] = -6$ and $[\a_{1}] =-8$ are irrelevant.
	\bigskip

	Going to dimensionless variables\footnote{Notation: the variables with a tilde are dimensionless.} and specialising the $Q$-functionals in \eqref{floweq1} to the optimised cutoff
	\beq
	\mathcal{R}_k(-\nabla^2) = (k^2 + \nabla^2)\theta(k^2+\nabla^2) \,,
	\eeq
	we can compute the beta functions ($\tilde\beta_\alpha= \partial_t\tilde\alpha$)
	\bea
	&&\tilde\beta_0=4 \text{$\tilde\alpha_0$}-\frac{\text{$\tilde\alpha_1$} }{6 \pi ^2}+\frac{\text{$\tilde\alpha _R$} g (\text{$\eta _\phi$}-4)}{\pi
		(1-2 \lambda )^2}\,, \\
	&&		\tilde\beta_{R} =\text{$\tilde\alpha_R$} \text{$\gamma_g$}+6\text{$\tilde\alpha_R$} - \frac{\text{$\tilde\alpha_1$} }{24 \pi ^2}+\frac{\text{$\tilde\alpha_R$} g (10
		\text{$\eta_\phi$} \lambda +4 \text{$\eta_\phi$}-30 \lambda -21)}{9 \pi  (2 \lambda -1)^3} \,, \\
	&& \tilde\beta_1 = \text{$\tilde\alpha_1$} \text{$\gamma_g$}+8 \text{$\tilde\alpha_1$}-\frac{\text{$\tilde\alpha_R$} g^2 (\text{$\eta_\phi$}-4)}{2 (2 \lambda
		-1)^3}+\frac{\text{$\tilde\alpha_1$} g (\text{$\eta_\phi$}-3)}{6 \pi  (1-2 \lambda )^2}-\frac{5 \text{$\tilde\alpha_1$} g }{24 \pi } \,.
	\eea
	where we have set the anomalous dimension $\hat \eta_k = 0$.
	
	We can write down the stability matrix as $S = \partial \tilde\beta_i/\partial \tilde\a_j$:
	\bea
	S=
	\left(
	\begin{array}{ccc}
		4$\qquad$ & \frac{g (\text{$\eta_\phi$}-4)}{\pi  (1-2 \lambda )^2} & -\frac{1}{6 \pi ^2} \\
		0$\qquad$&6+ \text{$\gamma_g$}+\frac{g (10 \text{$\eta_\phi$} \lambda +4 \text{$\eta_\phi$}-30 \lambda -21)}{9 \pi  (2 \lambda -1)^3} &
		-	\frac{1}{24 \pi ^2} \\
		0$\qquad$ & -\frac{g^2 (\text{$\eta_\phi$}-4)}{2 (2 \lambda -1)^3} &8+ \text{$\gamma_g$}+\frac{g (\text{$\eta_\phi$}-3)}{6 \pi  (1-2 \lambda
			)^2}-\frac{5 g }{24 \pi } \\
	\end{array}
	\right) \,.
	\eea

	When the couplings $\lambda$ and $g$ are $O(1)$ the dependence of the critical exponents  $\uptheta_1$ and $\uptheta_R$ on $\gamma_g$ is approximately linear up to corrections $10^{-7}$ provided we are sufficiently far from the pole $\lambda = 1/2$
	\beq
	\uptheta_i = \uptheta_i^{\textrm{canonical}} + a_i(\lambda, g)+ b_i(\lambda, g)\gamma_g + O(\gamma_g^2) \,.
	\label{gammag}
	\eeq
	Numerically $b_i(\lambda, g)\simeq- 1$.
	Therefore for every value of $g$ and $\lambda$ there is a critical value of $\gamma_g$ for which the quantum correction of the critical exponents vanishes. Increasing $\gamma_g$ always makes the critical exponents more irrelevant.
	
\begin{figure}[H]
		\centering
		\includegraphics[scale=0.55]{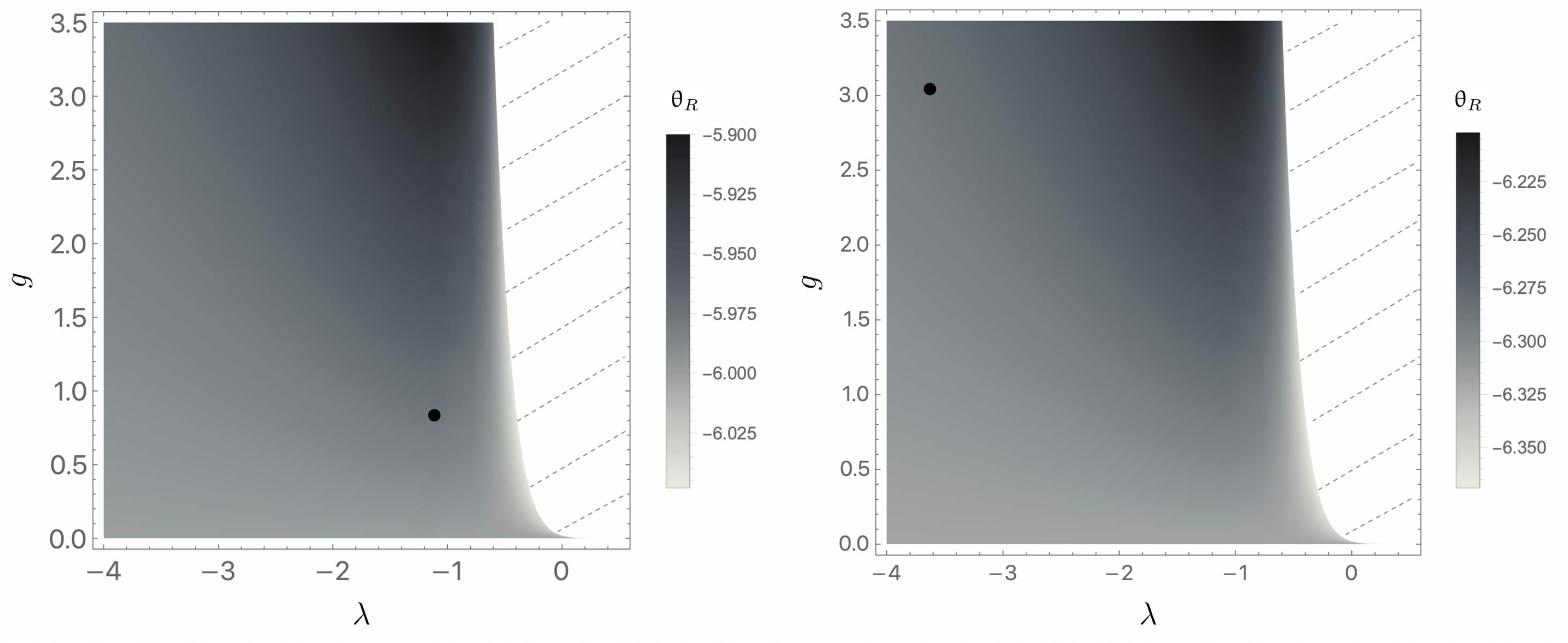}\quad 	\includegraphics[scale=0.55]{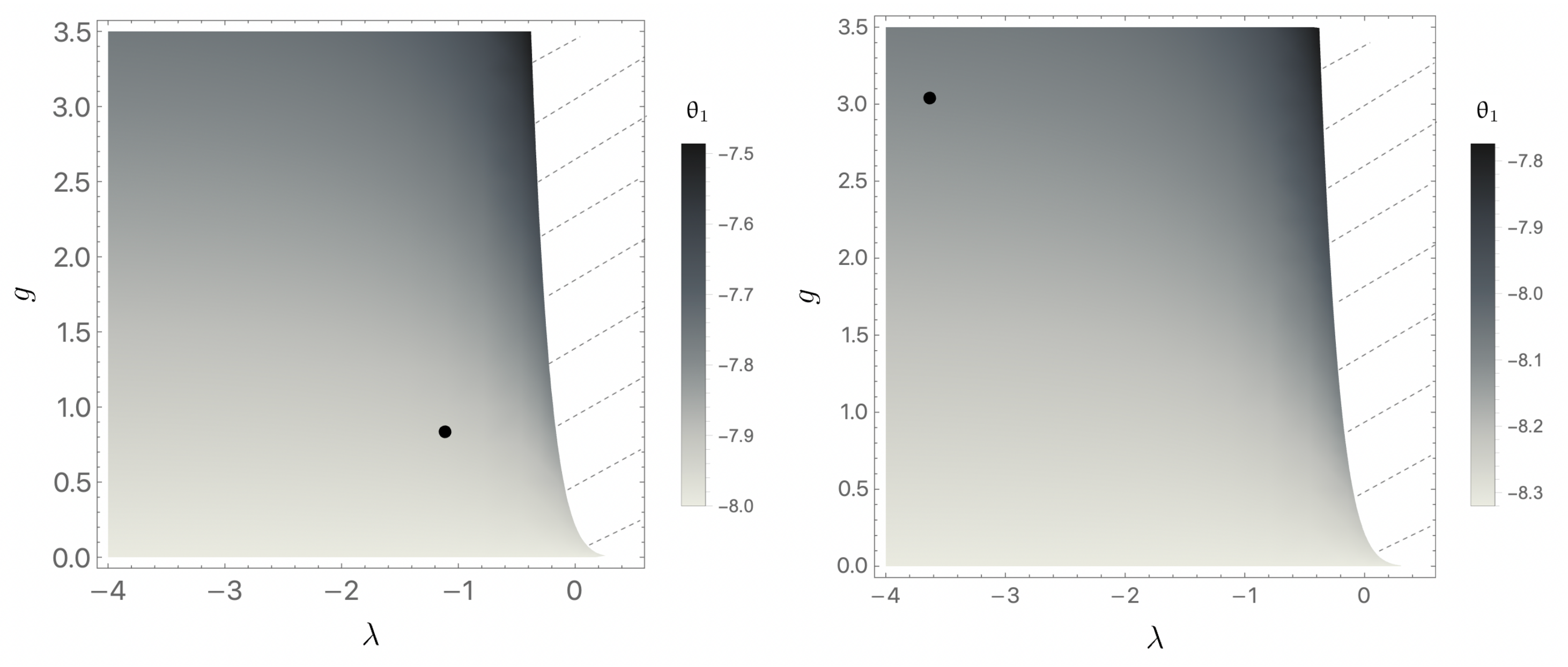}
		\caption{Plot of the critical exponents $\uptheta_R$ (up) and $\uptheta_1$ (down) of the stability matrix $S$ in the $\lambda - g$ - plane.  The plots on the left are for $\eta_N = -2$ and $\gamma_g$ = 0; the plots on the right are for $\eta_N = -2$ and $\gamma_g = 0.3197$ (these are the fixed points values which we discuss in Section V). The dashed region has been excluded from the plot because the critical exponents begin to diverge in this region due to the vicinity of a pole in the propagator. The black points represent the fixed point in the standard scheme with the Standard Model matter content. }\label{fig:density}
	\end{figure}

	In the plots in Figure \ref{fig:density} we are showing the critical exponents for $\uptheta_R$ and $\uptheta_1$ in the whole $g-\lambda$ plane.  We set  $\eta_N = -2$ and $\gamma_g$ = 0 for the plots on the left and  $\eta_N = -2$ and $\gamma_g = 0.3197$ for the plots on the right. These values correspond to the non-trivial fixed point values in two different renormalisation schemes (see Section \ref{sec:as}).
	We noticed that when $\gamma_g = 0$ the quantum corrections are negative, making the exponent more relevant, while when $\gamma_g = 0.3197$ the quantum corrections are positive, making the exponent more irrelevant. Note that along the axis $g=0$ the quantum corrections vanish if $\gamma_g = 0$ but remain for $\gamma_g = 0.3197$ since the scaling dimension of the metric remains anomalous.

	\section{Asymptotic safety}
	\label{sec:as}
	\subsection{Matter in Asymptotically safe gravity}
	\label{subsec:mattasymsafegra}
	Here we compute general expression for the scaling dimensions of relational observables which depend on the dimensionless Newton's constant $g$ and cosmological constant $\lambda$. The exact position of the fixed point depends on which matter fields we include.
	In general, we start considering for $N_S$ scalars, $N_D$ fermions, $N_V$ vectors and $N_D$ Dirac fermions minimally coupled to gravity. In order to write the beta functions, we have to define the cutoff: or a type II cutoff we choose a real function $\mathcal{R}_k(S^{(2)}_{\rm  matter})$ where  $S^{(2)}_{\rm  matter}$ is the inverse propagator of the matter field (for more details see e.g. \cite{perbook}).
	In $d=4$ the beta functions for the dimensionless Newton's coupling $g$ and the dimensionless cosmological constant $ \lambda$ with a type II cutoff for the matter fields become:
	\bea \label{betalambda}
	\partial_t\lambda =\lambda  (-2 \text{$\gamma_g$}+\text{$\eta_N$}-2)+\frac{g \left(\frac{5 (-2 \text{$\gamma_g$}+\text{$\eta_N$}-6)}{2 \lambda -1}+2 \text{$\eta_N$}-12 \text{$N_D$}+3 \text{$N_S$}+6
		\text{$N_V$}-24\right)}{12 \pi }\,, \\
	\label{betag}
	\partial_t g= g(\text{$\gamma_g$}+2) -
	\frac{g^2 \left(\frac{5 (2 \text{$\gamma_g$}-\text{$\eta_N$}+4)}{4 \lambda -2}+\frac{3 (2 \text{$\gamma_g$}-\text{$\eta_N$}+6)}{(1-2 \lambda )^2}-\frac{3 \text{$\eta_N$}}{2}+\text{$N_D$}-\text{$N_S$}+\text{$N_V$}+14\right)}{6 \pi } \,.
	\eea
	In the standard renormalisation scheme $\gamma_g = 0$.
	Since we neglect the anomalous dimensions of all the additional matter degrees of freedom ($\hat \eta_k=0$), we do not consider the effect of the anomalous dimension of the scalars $\hat X$'s.
	In this approximation the beta functions have a nontrivial fixed point ($ \partial_t g_\star =0 $ and   $\partial_t\lambda_\star=0$). Furthermore,  the Newton's constant is experimentally constrained to be positive, this puts a constraint on the matter content which possess a nontrivial fixed point.

	\subsubsection{Essential scheme}
	\label{subsec:essescheme}
	As it is explained in \cite{Baldazzi:2021ydj} and the Appendix \ref{app:essential}, the essential scheme allows the fields to be reparameterised along the RG flow using the RG kernel. At order $\partial^2$ the RG kernel of the metric and of the scalars $\hat X$ contain both one gamma function, $\gamma_g$ for the metric and minus one half times the anomalous dimension for the $\hat X$s. The anomalous dimension takes into account the normalisation of the kinetic term of the $\hat X$s, fixing their wave function renormalisation constant to one. The $\gamma_g$ can be used to fixed the vacuum energy $\tilde \rho\equiv \lambda/g$ to the value it takes at the Gaussian fixed point $(g,\lambda)=0$, which is given by
	\beq \label{tilderho_value}
	\tilde \rho \equiv \lambda/g =\frac{2 - 4 N_D + N_S + 2 N_V}{16 \pi}.
	\eeq
The anomalous dimension $\gamma_g$ is determined as a function of $g$ by setting $\partial_t  ( \lambda/g)  = 0$, in addition to \eqref{betalambda} and \eqref{betag}, with the vacuum energy fixed to the value \eq{tilderho_value} for all scales.  
	Indeed, at this order of the derivative expansion, the vacuum energy is the only inessential coupling, see \cite{Baldazzi:2021ydj}, and its flow can be fixed.
	Therefore, in the standard scheme, where we follow the flow of $(g,\lambda)$, we get that $\eta_N = \partial_t G/G \equiv \eta_\phi$ is equal to $-2$ at the Reuter fixed point. Using the essential scheme, instead, $\eta_\phi = \eta_N -2 \gamma_g$, where $\eta_N$ and $\gamma_g$ are both functions of $g$, because of the renormalisation condition that fixes the vacuum energy.
	
	\bigskip
	
	For the Standard Model and physically relevant modifications thereof, the fixed points have been computed in the standard and in the essential renormalisation scheme (see Table \ref{table1}). For the Standard Model, in order to investigate the cutoff dependence, we considered both a type II and a type I cutoff (with a type I cutoff we use the same profile function $\mathcal{R}_k$ but now with $-\nabla^2$ as its
	argument).Furthermore, we added an extra Scalar Field (SF), that could be a dark matter particle or an inflaton, for example, and the three neutrinos (3 $\nu$) to the Standard Model matter content.
	
	\begin{table}[H]
		\begin{center}
			\begin{minipage}{6cm}
				\begin{tabular}{|c|c|c|c|}\multicolumn{0}{c}{ 	Matter content}  &\multicolumn{0}{c}{ $N_S$}  &\multicolumn{0}{c}{ $N_D$}   & \multicolumn{0}{c}{ $N_V$}   \\   \hline
					\textbf{SM} (type II)&4&45/2&12\\ \hline
					\textbf{SM} (type I)&4&45/2&12\\ \hline
					\textbf{SM + SF}  (type II)&5&45/2&12\\ \hline
					\textbf{SM + 3 $\nu$ }(type II) &4&24&12\\ \hline
				\end{tabular}
			\end{minipage}
			\begin{minipage}[t]{4cm}
				\begin{tabular}{|c|c|c|c|}
					\multicolumn{1}{c}{\textbf{$\lambda_\star$}} &
					\multicolumn{1}{c}{\textbf{$g_\star$}} &
					\multicolumn{1}{c}{\textbf{$\eta_\phi$}} &
					\multicolumn{1}{c}{\textbf{$\gamma_g$}} \\ \hline
					-1.11626&0.834855& -2&0\\ \hline
					-3.58874&2.59505& -2&0\\ \hline
					-3.79874&2.77608& -2&0\\ \hline
					-4.83385&3.19355& -2&0\\ \hline
				\end{tabular}
			\end{minipage}
			\begin{minipage}[t]{5cm}
				\begin{tabular}{|c|c|c|c|}
					\multicolumn{1}{c}{\textbf{$\lambda_\star$}} &
					\multicolumn{1}{c}{\textbf{$g_\star$}} &
					\multicolumn{1}{c}{\textbf{$\eta_\phi$}} &
					\multicolumn{1}{c}{\textbf{$\gamma_g$}} \\ \hline
					-3.63186&3.04262&-2.63939&0.319697\\ \hline
					-1.11795&0.936571&-2.464896&0.232448\\ \hline
					-1.14584&0.959934&-2.470238&0.235119\\ \hline
					-1.07367&0.899477&-2.456218&0.228109\\ \hline
				\end{tabular}
			\end{minipage}
			\caption{Non-Gaussian fixed points. In the left table we show the matter content for each model. In the central and in the right table we show the fixed point values of the coupling constants and scaling dimensions in standard and in the essential renormalisation scheme respectively.}\label{table1}
		\end{center}
	\end{table}
	Let us remark that $\gamma_g$ is positive for the fixed points displayed in Table \ref{table1}, where Standard Model like  matter content were considered. This in contrast with the case of pure gravity where $\gamma_g$ is found to be negative at the Reuter fixed point \cite{Baldazzi:2021ydj}.

	\subsection{Results}
	We can now evaluate the scaling exponents 
	on the non-Gaussian fixed points which are displayed in Table \ref{table2}.
	\vspace{-0.7 cm}
	\begin{table}[H]
		\label{table2}
		\centering
		\begin{minipage}[t]{4cm}
			\begin{tabular}{|c|}\multicolumn{0}{c}{}   \\
				\multicolumn{0}{c}{Matter content}   \\ \hline
				\textbf{SM}	(type II)\\ \hline
				\textbf{SM}	(type I)\\ \hline
				\textbf{SM + SF} (type II)	\\ \hline
				\textbf{SM + 3 $\nu$} (type II) \\ \hline
			\end{tabular}
		\end{minipage}
		\begin{minipage}[t]{3,5cm}
			\begin{tabular}{|c|c|c|} \multicolumn{0}{c}{} \\
				\multicolumn{1}{c}{\textbf{$\uptheta_0$}} &
				\multicolumn{1}{c}{\textbf{$\uptheta_R$}} &
				\multicolumn{1}{c}{\textbf{$\uptheta_1$}}  \\ \hline
				-4&-5.97643&-7.92358\\ \hline
				-4&-5.97467&-7.8177\\ \hline
				-4&-5.97505&-7.80603\\ \hline
				-4&-5.98015&-7.78084\\ \hline
			\end{tabular}
		\end{minipage}
		\begin{minipage}[t]{3,5cm}
			\begin{tabular}{|c|c|c|} \multicolumn{0}{c}{} \\
				\multicolumn{1}{c}{\textbf{$\uptheta_0$}} &
				\multicolumn{1}{c}{\textbf{$\uptheta_R$}} &
				\multicolumn{1}{c}{\textbf{$\uptheta_1$}}  \\ \hline
				-4&-6.28653&-8.10472\\ \hline
				-4&-6.20271&-8.14459\\ \hline
				-4&-6.20467&-8.14593\\ \hline
				-4&-6.1996&-8.14236\\ \hline
			\end{tabular}
		\end{minipage}
		\caption{Critical exponents $\uptheta_i$ of the stability matrix. In the central and right table we show the values of the exponents for the standard and essential scheme respectively.}\label{table2}
	\end{table}
	The  scaling exponent of the field space volume term is not affected by any correction and keeps the canonical dimension 4. The other two  scaling exponents are affected by quantum corrections: $\uptheta_R$ and $\uptheta_1$ both become less irrelevant in the standard scheme, while they become more irrelevant in the essential one. This effect is due to the presence of $\gamma_g$ in the essential scheme, which pushes the  scaling exponents in being more irrelevant (see \eqref{gammag}).
	In any case, it is remarkable that the quantum corrections are very small in both schemes.
	
	\section{Higher order observables}
	\label{sec:general}
	In this section, we want to show how the procedure can be generalised to higher order observables. In fact, if we try to go to higher order in the derivative expansion, also terms at fourth order in derivative generated along the flow have to be taken into account.
	
	By plugging the observable \eqref{relactionorder2} in the composite operator flow equation, higher order relational observables are generated on rhs of the composite operator flow equation \eqref{flow_co}. For example we can consider higher powers of the relational metric. For this purpose  we consider the `building block matrix' $\M$
	\bea
	\M^{\hat \mu}_{\hat \nu}(x) =  g^{\mu \nu}(x) \partial_\mu X^{\hat \mu}(x) \partial_\nu X^{\hat \rho}(x)\hat \delta_{\hat \nu \hat \rho}
	\label{matrixM}
	\eea
	is related to the relational inverse metric by
	\beq
	\hat{\M}^{\hat{\mu}}_{\hat{\nu}}(\hat{x}) = \M^{\hat \mu}_{\hat \nu}(X(\hat{x}) )= \hat{g}^{\hat{\mu}\hat{\rho}}(\hat{x}) \delta_{\hat \nu \hat \rho}\,.
	\eeq
	For instance, at fourth order we come across terms like $\tr[\M^2]$ and $\tr[\M]^2$.
	However, at that order  we also encounter other terms. Imposing the $O(N)$ and shift symmetry constraints we have the following terms\footnote{It is important to emphasise that some terms which are absent in a $O(N)$ scalar model (see \cite{Peli:2020yiz} for a complete derivative expansion) in this context cannot be omitted since they do not represent  total derivatives. This is due to the different volume element $\tilde e$.}
	\bea
	\Gamma^{\rm rel.}_k = \int_x \tilde e && \left( \a_0 + \a_1 \tr [\M] + \a_R R +\a_{1,2}(\tr[\M])^2 +\a_2 \tr[\M^2 ]  \right.\nn &&\left. +\alpha_{box M}\Box (\tr \M)  +  \a_{\partial^4 X^4,1 } \delta_{\hat{\mu}\hat{\nu}}   (\Box \hat{X}^{\hat{\mu}})   (\Box \hat{X}^{\hat{\nu}})  +   \a_{\partial^4 X^4,2 } \delta_{\hat{\mu}\hat{\nu}}   ( \nabla_{\nu} \nabla_\mu \hat{X}^{\hat{\mu}})   ( \nabla^{\nu} \nabla^\mu \hat{X}^{\hat{\nu}})  \right.\nn   &&\left.+\a_{RM}R \;\tr [\M]+ \a_{Ricci} R^{ \mu \nu}  (\nabla_\nu X^{\hat{\nu}}) (\nabla_\mu X^{\hat{\mu}}) \delta_{\hat{\mu}\hat{\nu}} \right.\nn   &&\left.  + \alpha_{R^2}R^2+\alpha_{Ricci^2} R^{\mu \nu}R_{\mu \nu}+\alpha_{Riemann^2}R^{\mu \nu\alpha \beta}R_{\mu \nu\alpha \beta}  +   \alpha_{box R}  \Box R  \right) \,.
	\label{Gammarel4}
	\eea
	On the first line we have included the lower order terms already present in \eqref{ourgammarel} and the two terms which are quadratic in $\M$ which therefore involve four powers of the scalar fields $\hat{X}$. On the second line we include all terms with two powers of $\hat{X}$ and four derivatives, of which there are three linearly independent terms. In the last line we have terms which involve scalars which are fourth order in derivatives with no powers of the fields $\hat{X}$.
	
	Now we want to express the terms in the relational action as relational observables integrated over $\hat{x}$.
	One can introduce the connection
	\beq
	\hat \Gamma_{\hat \mu \hat \nu}^{\hat \rho} = - e_{\hat \mu}^{\mu}e_{\hat \nu }^\nu \nabla_{\mu}\partial_{\nu}\hat X ^{\hat \rho} \,.
	\eeq
	%
	%
	%
	%
	The expression of \eqref{Gammarel4} in terms of relational observable is given by
	\bea
	\Gamma^{\rm rel.}_k = \int_{\hat{x}} && \left( \a_0 + \a_1 \delta_{\hat{\mu}\hat{\nu}} \hat{g}^{\hat{\mu}\hat{\nu}} + \a_R \hat{R} +\a_{1,2}(\delta_{\hat{\mu}\hat{\nu}} \hat{g}^{\hat{\mu}\hat{\nu}})^2 +\a_2 \delta_{\hat{\mu}\hat{\nu}} \hat{g}^{\hat{\nu}\hat{\rho}}  \delta_{\hat{\rho}\hat{\lambda}} \hat{g}^{\hat{\lambda}\hat{\mu}}   \right.\nn &&\left. -2 \alpha_{box M}\Gamma^{\hat{\rho}}_{\hat{\rho} \hat{\mu}} \hat{g}^{\hat{\mu} \hat{\lambda}} \hat{\Gamma}_{\hat{\lambda} \hat{\sigma}}^{\hat{\tau}}  \hat{g}^{\hat{\sigma} \hat{\kappa}} \delta_{\hat{\kappa} \hat{\tau}} +  \a_{\partial^4 X^4,1 } \delta_{\hat{\mu}\hat{\nu}}   (\hat{g}^{\hat{\rho}\hat{\lambda}} \hat{\Gamma}^{\hat{\mu}}_{\hat{\rho} \hat{\lambda}})   (\hat{g}^{\hat{\tau}\hat{\sigma}} \hat{\Gamma}^{\hat{\nu}}_{\hat{\tau} \hat{\sigma}})  +   \a_{\partial^4 X^4,2 } \delta_{\hat{\mu}\hat{\nu}}   \hat{\Gamma}^{\hat{\mu}}_{\hat{\rho}\hat{\lambda}}  \hat{g}^{\hat{\rho} \hat{\tau}} \hat{g}^{\hat{\lambda}\hat{\sigma}}  \hat{\Gamma}^{\hat{\nu}}_{\hat{\tau}\hat{\sigma}}   \right.\nn   &&\left.+\a_{RM}\hat{R} \;\delta_{\hat{\mu}\hat{\nu}} \hat{g}^{\hat{\mu}\hat{\nu}}+ \a_{Ricci} \hat{R}^{ \hat{\mu} \hat{\nu}}  \delta_{\hat{\mu}\hat{\nu}} \right.\nn    &&\left.  + \alpha_{R^2}\widehat{R^2}+\alpha_{Ricci^2} \reallywidehat{R^{\mu \nu}R_{\mu \nu}}+\alpha_{Riemann^2}\reallywidehat{R^{\mu \nu \alpha \beta} R_{\mu \nu \alpha \beta} }  +   \alpha_{box R}  \widehat{\Box R}  \right)\,.
	\label{Gamma^rel_order4_RelOb}
	\eea
	Note that the canonical dimension of the last two terms of the first line and the terms in the second line is $-12$, the canonical dimension of the third line is $-10$ and the canonical dimension of the last line is $-8$.
	As we see, the terms in the second line are all expressed as different contractions of two powers of the relational inverse metric and two powers of the relational Christoffel symbol. While for the other terms the transition to relational observables is straight forward,  the first term on the second line deserves some explanation. First we note that expanding $\tilde e\Box (\tr \M) $ we  have
	$$
	\tilde e\Box (\tr \M) = 	\tilde e g^{\mu \nu}\Box (\partial_\mu \hat X^{\hat \mu}\partial_\nu \hat X^{\hat \nu}\hat \delta_{\hat\mu \hat \nu}) \,,
	$$
	which involves three derivatives acting on $\hat{X}$. However, we can instead integrate by parts to obtain
	$$
	\tilde e \Box (\tr \M) +\text{boundary terms}= -\nabla_\mu \tilde e \nabla^\mu (\tr \M) =-2\nabla_\rho \tilde e  g^{\mu \nu}\nabla^\rho( \partial_\mu \hat X^{\hat \mu}) \partial_\nu \hat X^{\hat \nu}\hat \delta_{\hat \mu \hat \nu} \,,
	$$
	which now involves only one or two derivatives acting on each $\hat{X}$. In terms of Christoffel symbols
	\beq
	\nabla_\mu \tilde e = - \tilde e e^{\hat\mu}_{\mu}\hat \Gamma^{\hat \rho}_{\hat \mu\hat \rho}, \qquad \nabla^\rho \partial_\nu \hat X ^{\hat \mu} =- e^{\rho}_{\hat \rho} e^{\hat \nu}_\nu \hat \Gamma^{\hat \mu}_{\hat \nu \hat \sigma} \hat g^{\hat \rho \hat \sigma} \,.
	\eeq
	It is then straightforward to see that this expression can be written in terms of the relational observables as in \eqref{Gamma^rel_order4_RelOb}.
	
	One last remark is in order: the scalar term $\tilde e \Box R$ could also be expressed as a relational observables linear in $\hat R$ and in the connection.	In fact, integrating by parts
	\bea
	\tilde e \Box R +\text{boundary terms}= - \nabla_{\mu} \tilde e \nabla^\mu R= \tilde e e^{\hat\mu}_{\mu}\hat \Gamma^{\hat \rho}_{\hat \mu\hat \rho}\nabla^\mu R \,.
	\eea

	\section{Conclusions}
	\label{sec:conclu}
	Seeking  a construction of observables within the asymptotic safety framework, we set up a general formalism for relational observables by introducing a set of four scalar fields. These fields represent the physical coordinate system upon which tensorial quantities can be evaluated and ``measured".
	
	Inspired by the well-established composite operator flow equation, we set up a formalism to evaluate the scaling of the relational observables. We furnish a definition for a relational action \eqref{relational_action} and  derive an equation describing its flow \eqref{flow_co}.
	Therefore, alongside the flowing effective average action, which interpolates between the microscopic action fixed point action and the full effective action, we also have access to flowing observables.
	
	Through this flow equation, we computed the scaling dimension of a selection of observables described in the relational action \eqref{relactionorder2}. This set of observables was chosen in order to self-consistently close the composite operator flow and to keep the truncation under control within a derivative expansion. It turned out that only tensorial quantities with upper indices satisfied this truncation requirement. In particular, we selected the inverse relational metric and the relational scalar curvature. On the other hand, tensorial quantities with lower indices, such as the relational metric, are not generated at any finite order in derivative expansion.
	\bigskip
	
	This formalism allows access to universal  critical exponents within any asymptotic safe quantum theory of gravity, beyond those that can be found by studying the flow of the effective action. This is important since it provides a new window through which we can compare different approaches to quantum gravity.
	For example the same exponents could be computed in Causal Dynamical Triangulations (see \cite{Loll:2019rdj}, \cite{Brunekreef:2020bwj} for curvature profiles, and \cite{Ambjorn:2021uge} for a recent analysis including four scalar fields), Tensor Field Theories and within a perturbative expansion around two dimensions.
	
It is important also to study the dependence of the critical exponents on the choice of regulator and on the choice of gauge.  Within simple approximation we might expect these dependencies to be strong and it will therefore be important to extend the current approximation by including more terms in the effective action beyond the Einstein-Hilbert action. We note however, that the value of critical exponent for the Newton's coupling is stable in the essential scheme between the Einstein-Hilbert approximation and the approximation where all terms with four derivatives are taken into account in the flow equation \cite{Baldazzi:2021orb}.  
	

	
	\bigskip
	
	Here we have applied our formalism within a simple approximation and taking the physical system to be a set of four scalar fields. This investigation can be extended in a number of ways. Firstly, as discussed in Section \ref{sec:general}, one could enlarge the set of observables in a consistent manner by going to higher orders in a derivative expansion. Furthermore, a more general set of observables is accessible both by including terms which break the shift and $O(N)$ symmetry and by including a non-constant source (see Section \ref{sec:rolesource}).  Finally, alternative coordinate systems can be constructed from other fields which allow access to an array of different relational observables \cite{Tambornino:2011vg}.

	\section*{Acknowledgements}
	We would like to thank Antônio Duarte Pereira, Martin Reuter, Carlo Pagani and Frank Saueressig for all the useful discussions.
	We are grateful to Roberto Percacci for his comments on the manuscript.
	

	\newpage
	\appendix
	\section{Essential scheme}\label{app:essential}
	In this section we review the essential RG approach \cite{Baldazzi:2021ydj} using the case of a single scalar field to avoid pure technicalities and then we present the generalisation for gravity developed in \cite{Baldazzi:2021orb}.
	In the standard approach the EAA obtains a dependence on the RG scale $k$ from one source: a momentum-dependent IR cutoff
	which  implements the coarse-graining procedure.
	In the essential scheme we introduce a second source of $k$ dependence, which takes into account the freedom to perform field reparameterisation along the flow parameterised by a $k$ diffeomorphism  $\hat{\phi}_k[\hat{\chi}]$  of configuration space which we integrate over in the functional integral. Explicitly, the EAA action $\Gamma_k[\phi]$ is defined by the functional integro-differential equation
	\bea \label{eq:defEAA}
	{\rm e}^{-\Gamma_k[\phi]} &:=&
	\int ({\rm{d}} \hat{\chi}) ~ {\rm e}^{-S[\hat{\chi}] + (\hat{\phi}_k[\hat{\chi}] - \phi) \cdot \frac{\delta}{\delta \phi} \Gamma_k[\phi]
		- \frac{1}{2}  ( \hat{\phi}_k[\hat{\chi}] - \phi) \cdot  \mathcal{R}_k \cdot (\hat{\phi}_k[\hat{\chi}] - \phi) }\,,
	\eea
	from which it follows that
	\beq
	\phi = \langle \hat{\phi}_k \rangle_{\phi,k} \,,
	\eeq
	and
	\beq
	\langle \hat{\mathcal{O}} \rangle_{\phi,k} :=   {\rm e}^{\Gamma_k[\phi]} \int ({\rm{d}} \hat{\chi}) ~ {\rm e}^{-S[\hat{\chi}] + (\hat{\phi}_k[\hat{\chi}] - \phi) \cdot \frac{\delta}{\delta \phi} \Gamma_k[\phi]
		- \frac{1}{2}  ( \hat{\phi}_k[\hat{\chi}] - \phi) \cdot  \mathcal{R}_k \cdot (\hat{\phi}_k[\hat{\chi}] - \phi) }  \hat{\mathcal{O}} [\hat{\chi}]\,,
	\eeq
	is the $\phi$ and $k$ dependent expectation value.
	In general, during the renormalisation process all operators compatible with the symmetries of the theory are generated but not all of them are associated to essential coupling, i.e., couplings that appears in physical quantities. Some of the couplings are inessentials meaning that they can be removed by field redefinitions.
	The utility of $\hat{\phi}_k[\hat{\chi}]$ is that we may choose to reparameterise the field to fix the values of inessential couplings.
	The generalised flow equation satisfied by $\Gamma_k[\phi]$ is given by \cite{Pawlowski:2005xe}
	\begin{equation} \label{scadimensionfull_flow}
		\left( \partial_t  +   \Psi_k[\phi] \cdot  \frac{\delta }{\delta \phi}  \right) \Gamma_k[\phi]
		= \frac{1}{2} \Tr \,  \mathcal{G}_k[\phi]  \left( \partial_t  + 2  \cdot \frac{\delta}{\delta \phi}  \Psi_k[\phi] \right)\cdot \mathcal{R}_k \,,
	\end{equation}
	where $t := \log(k/k_0)$, with $k_0$ some physical reference scale, under the trace appearing in the rhs
	\beq
	\mathcal{G}_k[\phi] :=  (\Gamma^{(2)}_k[\phi] + \mathcal{R}_k)^{-1}\,,
	\eeq
	is the IR regularised propagator, with $\Gamma^{(2)}_k[\phi] $ denoting the Hessian of the EAA with respect to the field $\phi(x)$, and
	\begin{equation}
		\Psi_k[\phi]  :=   \langle \partial_t \hat{\phi}_k[\hat{\chi}] \rangle_{\phi,k}\,
	\end{equation}
	is the RG kernel which takes into account the $k$-dependent field reparameterisations.
	The flow equation \eq{scadimensionfull_flow}  reduces to the standard flow for the EAA \cite{Wetterich:1992yh,Morris:1993qb} when $\Psi_k =0$.
	For quantum gravity the EAA is denoted $\Gamma_k[f;\bar{g}]$, where  $f= \{ g_{\mu\nu}, c^{\mu}, \bar{c}_{\mu} \}$ denotes the set of mean fields, where $g_{\mu\nu}$ is the (mean) metric, and $ c^{\mu}$ and  $\bar{c}_{\mu}$ are the (mean) anti-commuting ghost and anti-ghost. In addition to the mean fields, $\Gamma_k[f;\bar{g}]$ also depends on an auxiliary background metric $\bar{g}_{\mu\nu}$ in order to conserve background covariance. The EAA for gravity is defined analogously to the case of the scalar field \eq{eq:defEAA} though the functional integral
	\beq \label{EAA_def}
	{\rm{e}}^{- \Gamma_k[f,\bar{g}]} = \int \left({\rm{d}} \hat{\chi}\right)\, {\rm{e}}^{-S[\hat{\chi}; \bar{g} ]} {\rm{e}}^{(\hat{f}_k[\hat{\chi}] - f)\cdot \frac{\delta }{\delta f}  \Gamma_k[f;\bar{g}] }
	{\rm{e}}^{- \frac{1}{2} (\hat{f}_k[\hat{\chi}] - f)\cdot  \mathcal{R}_k[\bar{g}] \cdot  (\hat{f}_k[\hat{\chi}] - f) } \,,
	\eeq
	where $\hat{\chi}$ are a set of fields which parameterise the fields $\hat{f}_k[\hat{\chi}]=  \{ \hat{g}_{\mu\nu \, k}[\hat{\chi}], \hat{c}^{\mu}_{\,k}[\hat{\chi}], \hat{\bar{c}}_{\mu \, k}[\hat{\chi}] \}$ such that the latter defines a $k$-dependent diffeomorphism of the configuration space to itself.
	Similarly to the case of the scalar field, it follows from \eq{EAA_def} that
	\beq
	f = \langle \hat{f}_k \rangle_{f,k} \,,
	\eeq
	where the expectation value of any functional of the fields $\hat{\mathcal{O}}[\hat{\chi}]$ is defined by
	\beq
	\langle \hat{\mathcal{O}} \rangle_{f,k} :=  {\rm{e}}^{\Gamma_k[f,\bar{g}]} \int \left({\rm{d}} \hat{\chi}\right)\, {\rm{e}}^{-S[\hat{\chi}; \bar{g} ]} {\rm{e}}^{(\hat{f}_k[\hat{\chi}] - f)\cdot \frac{\delta }{\delta f}  \Gamma_k[f;\bar{g}] }
	{\rm{e}}^{- \frac{1}{2} (\hat{f}_k[\hat{\chi}] - f)\cdot  \mathcal{R}_k[\bar{g}] \cdot  (\hat{f}_k[\hat{\chi}] - f) }  \hat{\mathcal{O}}[\hat{\chi}] \,.
	\eeq
	The generalised flow equation for $\Gamma_k[f;\bar{g}]$ is given by
	\begin{equation} \label{dimensionfull_flow}
		\left( \partial_t  +   \Psi_k[f;\bar{g}] \cdot  \frac{\delta }{\delta f}  \right) \Gamma_k[f;\bar{g}]
		= \frac{1}{2} \STr \,  \mathcal{G}_k[f;\bar{g}]  \left( \partial_t  + 2  \cdot \frac{\delta}{\delta f}  \Psi_k[f;\bar{g}] \right)\cdot \mathcal{R}_k[\bar{g}] \,,
	\end{equation}
	where $f = \langle \hat{f} \rangle$ are the mean fields and $\mathcal{G}_k[f,\bar{g}]$ denotes the propagator
	\beq
	\mathcal{G}_k[f;\bar{g}] :=   \left(\frac{\delta}{\delta f } \Gamma_k[f;\bar{g}]  \frac{\overleftarrow{\delta}}{  \delta f }+ \mathcal{R}_k[\bar{g}] \right)^{-1} \,,
	\eeq
	with $\overleftarrow{\delta}$ signifying that the derivative acts to the left.
	The $\cdot$ implies a continuous matrix multiplication including sum over all field components and integration over spacetime. The $\STr$ denotes a supertrace in the same sense with a minus sign inserted for anti-commuting  fields.
	For gravity the RG kernel now has component for each field $\Psi_k = \{ \Psi_{\mu\nu}^g, \Psi^{c\mu}, \Psi^{\bar{c}}_\mu\}$ such that
	$
	\Psi_k  = \langle  \partial_t \hat{f}_k \rangle_{f,k}
	$\,.
	In the background field approximation, we choose $ \Psi^{c\mu} = 0 = \Psi^{\bar{c}}_\mu$, while we choose the RG kernel for the metric to be given by
	\beq \label{Psi_expansion}
	\Psi_{\mu\nu}^g[g] \equiv  \langle  \partial_t \hat{g}_{\mu\nu k} \rangle_{f,k}  = \gamma_g g_{\mu\nu}
	+ O(\partial^2) \,,
	\eeq
	where $\gamma_i$ with $i =\{ g, \ldots \}$ are the `gamma functions' which, along with the beta functions, are determined as functions of the couplings that appear in EAA.
	\section{Scalar tensor theory's flow equations}\label{app:scalartensor}
	In the standard approach the EAA that contains all terms at LPA' in the derivative expansion for a general scalar tensor theory\footnote{For a complete analysis about scalar tensor theories in asymptotic safety see \cite{Laporte:2021kyp}.} in Euclidean signature \cite{Steinwachs:2011zs} is
	\beq\label{eq:ansatzgamma}
	\Gamma_k = \int_x \sqrt{\det g}   \left[ \frac{\rho_k}{8\pi}  - Z_N R
	+ \frac{1}{2} Z_k\delta_{\hat\mu \hat\nu} \nabla_\mu \hat X^{\hat\mu} \nabla^\mu \hat X^{\hat\nu} \right]
	+   S_{\rm gf}[\bar{g},g] + S_{\rm gh}[g,\bar{g}, c,\bar{c}] \,,
	\eeq
	where $\rho_k = \Lambda_k/G_k$ and $Z_N= 1/(16\pi G_k)$.
	Using the essential scheme we can choose the RG kernel in order to impose the following renormalisation conditions along the RG flow
	\beq
	\rho_k = \rho_{GFP} = \frac{ 8\pi }{ d(4\pi)^{d/2} }  \int_0^\infty {\rm d} z  \,z^{\frac{d}{2}-1} \,\frac{\partial_t \mathcal{R}_k(z)}{z + \mathcal{R}_k(z)}   \,,
	\hspace{2cm} Z_k =1 \,.
	\eeq
	The first condition fixes the $k$-dependent vacuum energy to the value of the vacuum energy in the pure gravity GFP, while the second one is the standard renormalisation condition for the wave function renormalisation in order to canonically normalised the kinetic term.
	To achieve this, the RG kernels are
	\beq
	\Psi^{g}_{\mu\nu}  =  \gamma_g\, g_{\mu\nu} \,,
	\hspace{2cm}
	\Psi^{\hat\mu} =  -\frac{\hat \eta_k}{2} \hat X^{\hat\mu} \,,
	\eeq
	and in this way the EAA takes the following form
	\beq
	\Gamma_k = \int_x \sqrt{\det g}   \left[ \frac{\rho_{GFP}}{8\pi}  - Z_N R
	+ \frac{1}{2} \delta_{\hat\mu \hat\nu} \nabla_\mu \hat X^{\hat\mu} \nabla^\mu \hat X^{\hat\nu} \right]
	+   S_{\rm gf}[\bar{g},g] + S_{\rm gh}[g,\bar{g}, c,\bar{c}] \,.
	\eeq
	Using the previous RG kernels, on the rhs of \eqref{dimensionfull_flow} we get
	\bea
	&&\int_x
	\left(\frac{\delta \Gamma}{\delta \hat X^{\hat\mu}} \Psi^{\hat\mu}(x)
	+\frac{\delta \Gamma}{\delta g_{\mu\nu}(x)} \Psi^g(x)_{\mu\nu} \right)
	\\
	&=&\int_x \sqrt{\det g}  \left\{
	d Z_N \Lambda \gamma_g
	+\left[ -\frac{\hat \eta_k}{2} +
	\frac{d-2}{4}
	\gamma_g  \right]
	\delta_{\hat\mu \hat\nu}
	\nabla_\mu \hat X^{\hat\mu} \nabla^\mu \hat X^{\hat\nu}
	+ \frac{2-d}{2}
	Z_N \gamma_g \, R
	\right\} \,.
	\nonumber
	\eea
	In order to evaluate the lhs of \eqref{dimensionfull_flow}, we use the background field method with linear parameterisation
	\beq
	g_{\mu\nu} \to g_{\mu\nu} + \frac{1}{\sqrt{Z_N}} h_{\mu\nu} \,,
	\hspace{2cm}
	\hat X^{\hat \mu} \to \hat X^{\hat \mu} + \delta \hat X^{\hat \mu} \,,
	\eeq
	where both fluctuations around the background have the same dimensions.
	Because of the presence of $1/\sqrt{Z_N}$ in front of $h_{\mu\nu}$,
	the ghost are expanded in following way
	\beq
	C^{\mu} \to 
	Z_N^{-1/4} c^{\mu} \,,
	\hspace{2cm}
	\bar C_{\mu} \to 
	Z_N^{-1/4} \bar c_{\mu} \,,
	\eeq
	and we have
	\beq
	\Psi^{h}_{\mu\nu}  =  -\frac{1}{2}\left(\eta_N-2\gamma_g\right)\, h_{\mu\nu} \,,
	\hspace{2cm}
	\Psi^{c\mu} =  -\frac{\eta_N}{4} c^{\mu} \,,
	\hspace{2cm}
	\Psi^{\bar c}_{\mu} =  -\frac{\eta_N}{4} \bar c_{\mu} \,.
	\eeq
	Using de Donder gauge, the Hessian evaluated at zero value of the fluctuations reads
	\begin{align}
		&\frac{1}{\sqrt{{\rm det}g}}\frac{\delta^2\Gamma_k}{\delta h_{\mu\nu}\delta h_{\rho\sigma}}=
		K^{\mu\nu,\rho\sigma} \left( \Delta - 2\Lambda \right) + U^{\mu\nu \rho\sigma}
		+ \frac{1}{Z_N} S^{\mu\nu \rho \sigma} \,,
		\\
		&\frac{1}{\sqrt{{\rm det}g}}\frac{\delta^2\Gamma_k}{\delta \hat x^{\hat \mu} \delta \hat x^{\hat \nu} } =
		\delta_{\hat \mu \hat \nu}\Delta \,,
		\\
		&\frac{1}{\sqrt{{\rm det}g}}\frac{\delta^2\Gamma_k}{\delta \hat x^{\hat \mu} \delta h_{\mu\nu} } =
		- \frac{1}{\sqrt{Z_N}} K^{\mu\nu,\rho \sigma} \nabla_\rho \hat X_{\hat \mu} \nabla_\sigma\,,
		\\
		&\frac{1}{\sqrt{{\rm det}g}}\frac{\delta^2\Gamma_k}{\delta h_{\mu\nu} \delta \hat x^{\hat \mu} } =
		\frac{1}{\sqrt{Z_N}} K^{\mu\nu,\rho \sigma} \nabla_\rho \hat X_{\hat \mu} \nabla_\sigma
		+\frac{1}{2\sqrt{Z_N}} \left( g^{\mu\nu} \Delta +2 \nabla^\mu \nabla^\nu \right) \hat X_{\hat \mu} \,,
	\end{align}
	where
	\begin{align}
		& U^{\mu\nu}{}_{\rho\sigma} = K^{\mu\nu}{}_{\rho\sigma}R +\frac{1}{2} g^{\mu \nu} R_{\rho \sigma}+
		\frac{1}{2}R^{\mu \nu} g_{\rho \sigma}
		-\delta^{(\mu}_{(\rho}R^{\nu)}_{\sigma)}-R^{(\mu}{}_{(\rho}{}^{\nu)}{}_{\sigma)} \,,
		\\
		& S^{\mu\nu,\rho\sigma} =
		\left( -\frac{1}{2} K^{\mu\nu,\rho\sigma} g^{\a \b}
		+g^{\a ( \mu} \, \mathds{1}^{\nu) \b \rho\sigma}
		- \frac{1}{4} g^{\mu\nu} g^{\rho\a} g^{\sigma \b} - \frac{1}{4} g^{\rho\sigma} g^{\mu\a} g^{\nu \b}
		\right) \delta_{\hat\mu \hat\nu} \nabla_\a \hat X^{\hat\mu} \nabla_\b \hat X^{\hat\nu} \,,
		\\
		& K^{\mu\nu,\rho\sigma} = \frac{1}{4} \left( g^{\mu \rho}g^{\nu \sigma}+g^{\mu \sigma}g^{\nu \rho}-g^{\mu \nu}g^{\rho \sigma} \right) \,,
		\\
		&\left(K^{-1}\right)^{\mu\nu, \rho\sigma} =  g^{\mu \rho}g^{\nu \sigma}+g^{\mu \sigma}g^{\nu \rho}
		- \frac{2}{d-2}g^{\mu \nu}g^{\rho \sigma} \,.
	\end{align}
	Since we have mixed terms in the Hessian, we can proceed in different ways that differs by the renormalisation scheme procedure.
	\subsection{First possibility}
	\label{sec:firstpossibility}
	We are dealing with an Hessian of the form
	\bea
	\frac{1}{\sqrt{{\rm det}g}}\left(\Gamma^{(2)}\right)^{AB} = -\delta^{AB} g^{\mu\nu} \nabla_\mu\nabla_\nu - 2\Gamma^{AB\mu}\nabla_\mu + E^{AB}\,,
	\eea
	where
	\bea
	\Gamma_\sigma &=&\frac{1}{2\sqrt{Z_N}}K^{\mu\nu,\rho \sigma} \nabla_\rho \hat X_{\hat \mu}
	\begin{pmatrix}
		0 & -\mathds{1}  \\
		\mathds{1} & 0
	\end{pmatrix} \,,
	\\
	E &=&
	\begin{pmatrix}
		- 2\Lambda\, K^{\mu\nu,\rho\sigma} + U^{\mu\nu \rho\sigma}
		+ \frac{1}{Z_N} S^{\mu\nu \rho \sigma} & 0 \\
		0 & 0
	\end{pmatrix} \,.
	\eea
	Defining
		\begin{align}\label{eq:newcovder}
			\mathcal{D}^{AB}{}_{\sigma} &\equiv \delta^{AB} \nabla_\sigma + \Gamma^{AB}{}_\sigma \,,
	\end{align}
	such that this derivative applied to the fluctuations gives
	\begin{align}
		& \left( \mathcal{D} h \right)_{\alpha \beta,\mu} \equiv
		\nabla_\mu h_{\alpha \beta} + \Gamma{}_{\alpha \beta,\hat \mu , \mu} \delta \hat x^{\hat \mu}\,,
		\\
		& \left( \mathcal{D} \mathcal{D} h \right)_{\alpha \beta,\mu\nu} \equiv
		\nabla_\mu \nabla_\nu h_{\alpha \beta} + \Gamma{}_{\alpha \beta,\hat \mu , \mu} \Gamma{}^{\hat \mu, \gamma \delta}{}_{ \nu}  h_{\gamma \delta}
		+\nabla_\mu \left(\Gamma{}_{\alpha \beta,\hat \mu , \nu} \delta \hat x^{\hat \mu}\right)
		+ \Gamma{}_{\alpha \beta,\hat \mu , \mu} \nabla_\nu \delta \hat x^{\hat \mu}\,,
		\\
		& \left( \mathcal{D} \delta \hat x \right)^{\hat \mu}{}_{\mu} \equiv
		\nabla_\mu \delta \hat x^{\hat \mu} + \Gamma^{\hat \mu ,\alpha \beta}{}_\mu h_{\alpha \beta}\,,\\
		& \left( \mathcal{D} \mathcal{D} \delta \hat x \right)^{\hat \mu}{}_{\mu\nu} \equiv
		\nabla_\mu \nabla_\nu \delta \hat x^{\hat \mu}
		+ \Gamma^{\hat \mu}{}_{\alpha \beta,\mu} \Gamma^{\alpha \beta}{}_{\hat \nu , \nu} \delta \hat x^{\hat\nu}
		+ \nabla_\mu \left(\Gamma^{\hat \mu ,\alpha \beta}{}_\nu h_{\alpha \beta} \right)
		+ \Gamma^{\hat \mu ,\alpha \beta}{}_\mu \nabla_\nu h_{\alpha \beta}\,,
	\end{align}
	we can express the previous operator in the form
	\bea
	\mathcal{O}^{AB} &=& - g^{\mu\nu} \mathcal{D}^{AC}{}_\mu \mathcal{D}_C{}^B{}_\nu + \tilde{E}^{AB}\,,
	\\
	\tilde{E}^{AB} &=& E^{AB} + \nabla_\mu \Gamma^{AB\mu} + \Gamma^{AC\mu} \Gamma_C{}^{B\mu} \,,
	\\
	\Omega^{AB}_{\mu\nu} &\equiv& \left[ \mathcal{D}^{A}_{C}{}_{\mu} , \mathcal{D}^{CB}{}_{\nu} \right] = \nabla_\mu \Gamma^{AB}{}_\nu - \nabla_\nu \Gamma^{AB}{}_\mu
	+  \Gamma^{AC}{}_\mu \Gamma_C{}^B{}_\nu - \Gamma^{AC}{}_\nu \Gamma_C{}^B{}_\mu \,.
	\eea
	Note that $\tilde{E}$ and $\Omega$ have the following structures
	\bea
	\tilde{E} &=&
	\begin{pmatrix}
		- 2\Lambda\, K^{\mu\nu,\rho\sigma} + U^{\mu\nu \rho\sigma}
		+ \frac{1}{Z_N} S^{\mu\nu \rho \sigma} + \left(\Gamma^2_{11}\right)^{\mu\nu \rho \sigma}&\hspace{1cm} \nabla_\alpha \Gamma^{\hat\nu\rho\sigma,\alpha} \\
		\nabla_\alpha \Gamma^{\mu\nu\hat\mu,\alpha} & \left(\Gamma^2_{22}\right)^{\hat\mu\hat\nu}
	\end{pmatrix}\\
	\Omega_{\alpha\beta} &=&
	\begin{pmatrix}
		\Gamma^{\mu\nu\hat\mu}{}_{\alpha} \Gamma_{\hat\mu}{}^{\rho\sigma}{}_{\beta} - \Gamma^{\mu\nu\hat\mu}{}_{\beta}\Gamma_{\hat\mu}{}^{\rho\sigma}{}_{\alpha} & \nabla_\alpha \Gamma^{\hat\nu\rho\sigma}{}_{\beta}-\nabla_\beta \Gamma^{\hat\nu\rho\sigma}{}_{\alpha} \\
		\nabla_\alpha \Gamma^{\mu\nu\hat\mu}{}_{\beta} - \nabla_\beta \Gamma^{\mu\nu\hat\mu}{}_{\alpha} &\hspace{1cm} \Gamma^{\hat\mu\mu\nu}{}_{\alpha} \Gamma_{\mu\nu}{}^{\hat\nu}{}_{\beta} - \Gamma^{\hat\mu\mu\nu}{}_{\beta} \Gamma_{\mu\nu}{}^{\hat\nu}{}_{\alpha}
	\end{pmatrix}
	\eea
	where
	\bea
	\left(\Gamma^2_{11}\right)^{\mu\nu}{}_{\alpha\beta} &\equiv& \Gamma^{AC\mu} \Gamma_{CB}{}^{\mu}\Big|_{A=\mu\nu,B=\alpha\beta}
	= - \frac{1}{4Z_N} K^{\mu\nu,\rho\sigma}
	K_{\alpha\beta}{}^\gamma{}_\sigma \nabla_\rho \hat X_{\hat\mu} \nabla_\gamma \hat X^{\hat\mu} \,,
	\\
	\left(\Gamma^2_{22}\right)^{\hat\mu}{}_{\hat\nu} &\equiv& \Gamma^{AC\mu} \Gamma_{CB}{}^{\mu}\Big|_{A=\hat\mu,B=\hat\nu}
	= -\frac{1}{4Z_N} K^{\mu\nu,\rho\sigma}
	K_{\mu\nu}{}^\alpha{}_\sigma \nabla_\rho \hat X^{\hat\mu} \nabla_\alpha \hat X_{\hat\nu} \,.
	\eea
	It is important to stress out that the mixed terms of $\Omega_{\alpha\beta}$ are total derivatives.\\We then insert the regulator in such a way that
	$$ - g^{\mu\nu} \mathcal{D}^{AC}{}_\mu \mathcal{D}_C{}^B{}_\nu \to P_k{}^{AB} \equiv - g^{\mu\nu} \mathcal{D}^{AC}{}_\mu \mathcal{D}_C{}^B{}_\nu + \mathcal{R}_k{}^{AB} \,.
	$$
	Thus, the propagator has the following form
	\begin{align}\label{eq:proptot}
		\mathcal{G}_k = \frac{1}{\sqrt{{\rm det}g}}
		\begin{pmatrix}
			\mathcal{G}_{gg} & 0  \\
			0 & \mathcal{G}_{XX}
		\end{pmatrix} \,,
	\end{align}
	where
	\begin{align}\label{eq:propgrav}
		\left(\mathcal{G}_{gg}\right)_{\mu\nu\rho\sigma} &= \left(K^{-1}\right)_{\mu\nu\rho\sigma}\frac{1}{P_k-2\Lambda}
		- \frac{1}{P_k-2\Lambda}
		\left[K^{-1}\cdot\left( U + \frac{1}{Z_N} S + \Gamma_{11}^2\right)\cdot K^{-1}\right]_{\mu\nu\rho\sigma} \frac{1}{P_k-2\Lambda}\,,
		\\
		\left(\mathcal{G}_{XX}\right)_{\hat\mu\hat\nu} &= \delta_{\hat\mu\hat\nu} \frac{1}{P_k}
		- \frac{1}{P_k}
		\left(\Gamma_{22}^2\right)_{\hat\mu\hat\nu} \frac{1}{P_k}\,,
		\label{eq:propsca}
	\end{align}
	modulo four derivative terms, coming from squares of the curvature and derivatives of $\hat X$.
	The lhs of \eqref{dimensionfull_flow} is composed by three traces, which read the following
	\bea
	\mathcal{T}_{hh} &=&
	\frac{1}{2} \Tr \left[\frac{1}{P_k-2\Lambda}
	- \frac{K^{-1}}{P_k-2\Lambda}
	\left( U + \frac{1}{Z_N} S + \Gamma_{11}^2\right)\frac{1}{P_k-2\Lambda}
	\right]\,K^{-1}\,\left(\partial_t -\eta_N +2\gamma_g\right) \mathcal{R}^{hh}_k \,,
	\\
	\mathcal{T}_{xx} &=&
	\frac{1}{2} \Tr \left[\frac{1}{P_k}
	- \frac{1}{P_k}
	\Gamma_{22}^2 \frac{1}{P_k}
	\right]\,\left(\partial_t - \hat \eta_k\right) \mathcal{R}^{xx}_k \,,
	\\
	\mathcal{T}_{\bar c c} &=&
	- \Tr \left[\frac{1}{P_k}
	+ Ricci \frac{1}{P_k^2}
	\right]\, \left(\partial_t -\frac{\eta_N}{2}\right) \mathcal{R}^{\bar c c}_k \,,
	\eea
	and the regulator is chosen in the following way
	\footnote{Note that the factor $\sqrt{{\rm det}g}$ in the propagator and in the regulator cancels when they are multiplied.}
	\begin{align}\label{eq:regchoice}
		\left(\mathcal{R}_k^{hh}\right)^{\mu\nu,\alpha\beta} = \sqrt{{\rm det}g}\,K^{\mu\nu,\alpha\beta} \,\mathcal{R}_k \,,
		\hspace{1cm} \left(\mathcal{R}_k^{xx}\right)_\mu^\nu = \sqrt{{\rm det}g}\,\hat\delta_{\hat\mu}^{\hat\nu} \mathcal{R}_k \,,
		\hspace{1cm} \left(\mathcal{R}_k^{\bar c c}\right)_\mu^\nu = \sqrt{{\rm det}g}\, \delta_\mu^\nu \mathcal{R}_k \,.
	\end{align}
	Defining $\eta_N = -\partial_t \ln Z_N$ , the traces for a generic regulator $\mathcal{R}_k$ and generic dimensions are
	\bea
	\mathcal{T}_{hh} &=&
	\frac{1}{2(4\pi)^{d/2}} \int {\rm d}^d\,x  \sqrt{{\rm det}g}\,\Big\{
	\frac{d(d+1)}{2} Q_{d/2}\left[ \frac{(\partial_t-\eta_N +2 \gamma_g)\mathcal{R}_k}{P_k -2\Lambda} \right] + \frac{d(d+1)}{12}\,R
	\,Q_{d/2-1}\left[ \frac{(\partial_t-\eta_N +2 \gamma_g)\mathcal{R}_k}{P_k -2\Lambda} \right]
	\nonumber\\
	&& \hspace{2cm}
	- \left( \frac{d(d-1)}{2}\,R -\frac{1}{Z_N}\left(\frac{(d+1)(d-4)}{4} + \frac{d}{16} \right)
	\nabla_\mu \hat X_{\hat\mu} \nabla^\mu \hat X^{\hat\mu}\right)
	Q_{d/2}\left[ \frac{(\partial_t-\eta_N +2 \gamma_g)\mathcal{R}_k}{\left(P_k -2\Lambda\right)^2} \right]
	\Big\} \,,
	\nonumber\\
	\mathcal{T}_{xx} &=&
	\frac{1}{2(4\pi)^{d/2}} \int {\rm d}^d\,x \sqrt{{\rm det}g}\,\Big\{
	N_S \,Q_{d/2}\left[ \frac{(\partial_t-\hat \eta_k)\mathcal{R}_k}{P_k} \right]
	+ \frac{N_S}{6}\,R\,Q_{d/2-1}\left[ \frac{(\partial_t-\hat \eta_k)\mathcal{R}_k}{P_k} \right]
	\nonumber\\
	&& \hspace{3cm}
	+ \frac{1}{Z_N}\frac{3d-2}{64} \nabla_\mu \hat X_{\hat\mu} \nabla^\mu \hat X^{\hat\mu} \,Q_{d/2}\left[ \frac{(\partial_t-\hat \eta_k)\mathcal{R}_k}{P_k^2} \right]
	\Big\}\,,
	\nonumber\\
	\mathcal{T}_{\bar c c} &=&
	-\frac{1}{(4\pi)^{d/2}} \int {\rm d}^d\,x  \sqrt{{\rm det}g}\,\Big\{
	d \,Q_{d/2}\left[ \frac{\left(\partial_t-\frac{\eta_N}{2}\right)\mathcal{R}_k}{P_k} \right]
	+ \frac{d}{6}R\,Q_{d/2-1}\left[ \frac{\left(\partial_t-\frac{\eta_N}{2}\right)\mathcal{R}_k}{P_k} \right]
	+ R \,Q_{d/2}\left[ \frac{\left(\partial_t-\frac{\eta_N}{2}\right)\mathcal{R}_k}{P_k^2} \right]
	\Big\}\,.
	\nonumber
	\eea
	Putting everything together, the flow equations are
	\bea
	\left( 2\partial_t + d  \gamma_g \right) Z_N \Lambda
	&=&
	\frac{1}{2(4\pi)^{d/2}}\,\Big\{
	\frac{d(d+1)}{2} Q_{d/2}\left[ \frac{(\partial_t-\eta_N +2 \gamma_g)\mathcal{R}_k}{P_k -2\Lambda} \right]
	+ N_S \,Q_{d/2}\left[ \frac{(\partial_t-\hat \eta_k)\mathcal{R}_k}{P_k} \right]
	\nonumber\\
	&&\hspace{2cm}- 2 d \,Q_{d/2}\left[ \frac{\left(\partial_t-\frac{\eta_N}{2}\right)\mathcal{R}_k}{P_k} \right]
	\Big\}\,,
	\\
	- \left( \partial_t + \frac{d-2}{2} \gamma_g \right)\,Z_N
	&=&
	\frac{1}{2(4\pi)^{d/2}} \Big\{
	\frac{d(d+1)}{12}\,
	\,Q_{d/2-1}\left[ \frac{(\partial_t-\eta_N +2 \gamma_g)\mathcal{R}_k}{P_k -2\Lambda} \right]
	- \frac{d(d-1)}{2}
	Q_{d/2}\left[ \frac{(\partial_t-\eta_N +2 \gamma_g)\mathcal{R}_k}{\left(P_k -2\Lambda\right)^2} \right]
	\nonumber\\
	&& \hspace{0cm}
	+\frac{N_S}{6}\,Q_{d/2-1}\left[ \frac{(\partial_t-\hat \eta_k)\mathcal{R}_k}{P_k} \right]
	- \frac{d}{3}\,Q_{d/2-1}\left[ \frac{\left(\partial_t-\frac{\eta_N}{2}\right)\mathcal{R}_k}{P_k} \right]
	-2\,Q_{d/2}\left[ \frac{\left(\partial_t-\frac{\eta_N}{2}\right)\mathcal{R}_k}{P_k^2} \right]
	\Big\}\,,
	\\
	-\frac{\hat \eta_k}{2} +\frac{d-2}{4}\gamma_g
	&=& \frac{1}{2(4\pi)^{d/2}} \frac{1}{Z_N}\left\{
	\left(\frac{(d+1)(d-4)}{4} + \frac{d}{16} \right)
	Q_{d/2}\left[ \frac{(\partial_t-\eta_N +2 \gamma_g)\mathcal{R}_k}{\left(P_k -2\Lambda\right)^2} \right]
	\right.
	\nonumber\\
	&& \hspace{3cm}\left.
	+\frac{3d-2}{64} \,Q_{d/2}\left[ \frac{(\partial_t-\hat \eta_k)\mathcal{R}_k}{P_k^2} \right] \right\}\,.
	\eea
	\subsection{Second possibility}
	\label{app:secposs}
	We insert the regulator in such a way that $\Delta \to P_k \equiv \Delta + \mathcal{R}_k(\Delta)$, i.e., Equation~\eqref{eq:regchoice}, and then we calculate the inverse regularised propagator at first order in the curvature and second order in derivative of the scalars.
	The regularised Hessian takes the form
	\bea
	\frac{1}{\sqrt{{\rm det}g}}\left(\Gamma_k^{(2)} + \mathcal{R}_k \right) =\left(
	\begin{array}{cc}
		K_{\mu\nu}{}^{\rho \sigma} \left(P_k-2\Lambda\right)
		+ U_{\mu\nu}{}^{\rho \sigma} + \frac{1}{Z_N} S_{\mu\nu}{}^{\rho \sigma}
		& - \frac{1}{\sqrt{Z_N}} K_{\mu\nu}{}^{\gamma \delta} \nabla_\gamma \hat X_{\hat \mu} \nabla_\delta
		\\
		\frac{1}{\sqrt{Z_N}} K^{\rho \sigma}{}^{\gamma \delta} \nabla_\gamma \hat X_{\hat \mu} \nabla_\delta
		&  \hat\delta_{\hat \mu \hat \nu} P_k  \\
	\end{array}
	\right) \,,
	\eea
	and its inverse is
	\begin{subequations}
		\begin{align}
			&\sqrt{{\rm det}g}\left[\left(\Gamma_k^{(2)} + \mathcal{R}_k \right)^{-1}_{11} \right]_{\rho\sigma}{}^{\alpha \beta}=
			\frac{1}{P_k-2\Lambda} \left( K^{-1} \right)_{\rho \sigma}{}^{\alpha \beta}
			- \frac{1}{\left(P_k-2\Lambda\right)^2} \left( K^{-1} \right)_{\rho\sigma}{}^{\gamma\delta}
			\left(U+\frac{1}{Z_N}S\right)_{\gamma\delta}{}^{\epsilon\eta}
			\left( K^{-1}\right)_{\epsilon \eta}{}^{\alpha \beta}
			\\
			& \hspace{3cm}- \frac{1}{Z_N}\frac{1}{P_k\left(P_k-2\Lambda\right)^2}
			\left( K^{-1}\right)_{\rho \sigma}{}^{\gamma \delta}
			K_{\gamma \delta}{}^{\epsilon \eta} \nabla_\epsilon \hat X_{\hat \mu}
			K^{\tau \phi}{}^{\lambda \omega} \nabla_\lambda \hat X^{\hat \mu}
			\left(K^{-1}\right)_{\tau \phi}{}^{\alpha \beta} \nabla_\eta \nabla_\omega \,,
			\nonumber\\
			& \sqrt{{\rm det}g}\left[\left(\Gamma_k^{(2)} + \mathcal{R}_k \right)^{-1}_{12} \right]_{\rho\sigma\hat\mu } =
			\frac{1}{\sqrt{Z_N}}\frac{1}{P_k \left( P_k -2\Lambda \right)}
			K_{\rho \sigma}{}^{\lambda \omega} \nabla_\lambda \hat X_{\hat \mu}
			\left(K^{-1}\right)_{\rho \sigma}{}^{\alpha \beta} \nabla_\omega \,,
			\\
			& \sqrt{{\rm det}g}\left[\left(\Gamma_k^{(2)} + \mathcal{R}_k \right)^{-1}_{21} \right]^{\alpha \beta}{}_{\hat\mu}=
			-\frac{1}{\sqrt{Z_N}}\frac{1}{P_k \left( P_k -2\Lambda \right)}
			K^{\rho \sigma}{}^{\lambda \omega} \nabla_\lambda \hat X_{\hat \mu}
			\left(K^{-1}\right)_{\rho \sigma}{}^{\alpha \beta} \nabla_\omega \,,
			\\
			& \sqrt{{\rm det}g}\left[\left(\Gamma_k^{(2)} + \mathcal{R}_k \right)^{-1}_{22}\right]_{\hat\mu \hat\nu} =
			\hat \delta_{\hat \mu \hat \nu}\frac{1}{P_k}
			-\frac{1}{P_k^2 \left( P_k -2\Lambda \right)}
			K_{\alpha \beta}{}^{\epsilon \eta} \nabla_\epsilon \hat X_{\hat \mu}
			K^{\tau \phi}{}^{\lambda \omega} \nabla_\lambda \hat X_{\hat \nu}
			\left(K^{-1}\right)_{\tau \phi}{}^{\alpha \beta} \nabla_\eta \nabla_\omega \,.
		\end{align}
	\end{subequations}
	Then the traces read
	\bea
	\mathcal{T}_{hh} &=&
	\frac{1}{2(4\pi)^{d/2}} \int {\rm d}^d\,x  \sqrt{{\rm det}g}\,\left\{
	\frac{d(d+1)}{2} Q_{d/2}\left[ \frac{(\partial_t-\eta_N +2 \gamma_g)\mathcal{R}_k}{P_k -2\Lambda} \right] + \frac{d(d+1)}{12}\,R
	\,Q_{d/2-1}\left[ \frac{(\partial_t-\eta_N +2 \gamma_g)\mathcal{R}_k}{P_k -2\Lambda} \right]
	\right.\nonumber\\
	&& \left.\hspace{2cm}
	- \left( \frac{d(d-1)}{2}\,R -\frac{1}{Z_N}\frac{(d+1)(d-4)}{4}
	\nabla_\mu \hat X_{\hat\mu} \nabla^\mu \hat X^{\hat\mu}\right)
	Q_{d/2}\left[ \frac{(\partial_t-\eta_N +2 \gamma_g)\mathcal{R}_k}{\left(P_k -2\Lambda\right)^2} \right]
	\right.\nonumber\\
	&& \left.\hspace{2cm}
	+\frac{1}{Z_N} \left(\frac{d+1}{2} - \frac{1}{d-2} \right)
	\nabla_\mu \hat X_{\hat\mu} \nabla^\mu \hat X^{\hat\mu}
	Q_{d/2+1}\left[ \frac{(\partial_t-\eta_N +2 \gamma_g)\mathcal{R}_k}{P_k\left(P_k -2\Lambda\right)^2} \right]
	\right\} \,,
	\nonumber\\
	\mathcal{T}_{xx} &=&
	\frac{1}{2(4\pi)^{d/2}} \int {\rm d}^d\,x \sqrt{{\rm det}g}\,\left\{
	N_S \,Q_{d/2}\left[ \frac{(\partial_t-\hat \eta_k)\mathcal{R}_k}{P_k} \right]
	+ \frac{N_S}{6}\,R\,Q_{d/2-1}\left[ \frac{(\partial_t-\hat \eta_k)\mathcal{R}_k}{P_k} \right]
	\right.\nonumber\\
	&& \hspace{3cm} \left.
	+ \frac{1}{Z_N} \frac{d}{8} \nabla_\mu \hat X_{\hat\mu} \nabla^\mu \hat X^{\hat\mu} \,Q_{d/2+1}\left[ \frac{(\partial_t-\hat \eta_k)\mathcal{R}_k}{P_k^2(P_k-2\Lambda)} \right]
	\right\} \,,
	\nonumber\\
	\mathcal{T}_{\bar c c} &=&
	-\frac{1}{(4\pi)^{d/2}} \int {\rm d}^d\,x  \sqrt{{\rm det}g}\,\left\{
	d \,Q_{d/2}\left[ \frac{\left(\partial_t-\frac{\eta_N}{2}\right)\mathcal{R}_k}{P_k} \right]
	+ \frac{d}{6}R\,Q_{d/2-1}\left[ \frac{\left(\partial_t-\frac{\eta_N}{2}\right)\mathcal{R}_k}{P_k} \right]
	+ R \,Q_{d/2}\left[ \frac{\left(\partial_t-\frac{\eta_N}{2}\right)\mathcal{R}_k}{P_k^2} \right]
	\right\} \,.
	\nonumber
	\eea
	\section{Hessian of the observable}\label{app:hessian}
	In this section, we present the details of the calculation of the Hessian of the observable
	\begin{align}\label{eq:observablechoosen}
		\Gamma^{\rm rel.}_k = \int_x \tilde e \left( \alpha_0 + \alpha_R \,R + \alpha_1\,{\rm tr}\left[\M\right] \right)\,.
	\end{align}
	The first and second variations read
	\begin{align}
		\delta \tilde e &= \tilde e \, e^\mu_{\hat\mu} \partial_\mu \delta \hat X^{\hat\mu}\,,\\
		\delta^2 \tilde e &= 2 \tilde e \, e^{[\mu}_{\hat\mu}e^{\nu]}_{\hat\nu}
		\partial_\mu \delta \hat X^{\hat\mu}\partial_\nu \delta \hat X^{\hat\nu}\,,\\
		\delta {\rm tr}\M &=  2g^{\mu \nu} \hat \delta_{\hat \mu \hat \nu}\partial_\mu\delta \hat X^{\hat \mu}\partial_\nu \hat X^{\hat \nu} + \delta g^{\mu \nu} \hat \delta_{\hat \mu \hat \nu}\partial_\mu \hat X^{\hat \mu}\partial_\nu \hat X^{\hat \nu} \,,\\
		\delta^2 {\rm tr}\M &=  2 g^{\mu \nu} \hat \delta_{\hat \mu \hat \nu} \partial_\mu\delta \hat X^{\hat \mu}\partial_\nu \delta \hat X^{\hat \nu} + 4 \delta g^{\mu \nu} \hat \delta_{\hat \mu \hat \nu}\partial_\mu\delta  \hat X^{\hat \mu}\partial_\nu \hat X^{\hat \nu} + \delta^2 g^{\mu \nu}  \hat \delta_{\hat \mu \hat \nu}\partial_\mu \hat X^{\hat \mu}\partial_\nu \hat X^{\hat \nu} \,.
	\end{align}
	Since it holds
	\begin{align}
		\nabla_\mu \tilde e = \tilde e e^\alpha_{\hat \mu} \nabla_\alpha e^{\hat\mu}_\mu =- \tilde e e^{\hat\mu}_\mu \nabla_\alpha e^\alpha_{\hat\mu}
		\implies
		\nabla_\mu \left( \tilde e \,e^{\mu}_{\hat\mu} \right)=0 \,,
	\end{align}
	then
	\begin{align}\label{eq:idtildeeapp}
		\int_x \delta \tilde e = \int_x \partial_\mu \left( \tilde e \,e^{\mu}_{\hat\mu} \delta \hat X^{\hat\mu}\right)\,.
	\end{align}
	Thus, the variation of the term coupled to $\alpha_0$ is identically zero: this is a consequence of the fact that this term is a total derivative and so it behaves like a topological term.
	Moreover, the following identities hold
	\begin{align}\label{eq:firstvarapp}
		\delta \int_x \tilde e \, S &= \int_x \tilde e\left( \delta S - e^\mu_{\hat\mu} \delta \hat X^{\hat\mu} \partial_\mu S \right)\,,\\
		\label{eq:secondvarapp}
		\delta^2 \int_x \tilde e \, S &= \int_x \tilde e\left( \delta^2 S -2 e^\mu_{\hat\mu} \delta \hat X^{\hat\mu} \partial_\mu \delta S + 2 e^{[\mu}_{\hat\mu}e^{\nu]}_{\hat\nu} \partial_\mu \delta \hat X^{\hat\mu}\partial_\nu \delta \hat X^{\hat\nu}\,S \right)\,.
	\end{align}
	
	Using the essential scheme on the lhs we have the following additional contribution\footnote{Note that using Equations~\eqref{eq:firstvarapp} and \eqref{eq:idtildeeapp} the variation of $\tilde e$ gives no contribution.}
	\begin{align}
		\int_x \frac{\delta \Gamma^{\rm rel.}}{\delta g_{\mu\nu}}\, \Psi^g_{\mu\nu}+ \frac{\delta \Gamma^{\rm rel.}}{\delta \hat X^{\hat\mu}}\, \Psi^x_{\hat\mu}
		= \int_x \tilde e\,\left\{ -\gamma_g \, \alpha_R\,R
		-\left( \gamma_g + \hat\eta_k \right) \alpha_1\,{\rm tr}\M\right\}\,.
	\end{align}
	
	The Hessian of the term associated to $\alpha_1$ is
	\begin{align}
		\frac{1}{Z_N}g_{\rho\alpha}g_{\sigma\beta}\frac{\delta^2}{\delta g_{\mu\nu}\delta g_{\alpha\beta}}\int_x \tilde e \, {\rm tr} \M &=
		2\tilde e\,\delta^{(\mu}_{(\rho}g^{\nu)\alpha}\delta^{\beta}_{\sigma)} \hat \delta_{\hat\mu \hat\nu}\partial_\alpha \hat X^{\hat\mu} \partial_\beta \hat X^{\hat\nu} \,,
		\\
		\frac{\delta^2}{\delta \hat X^{\hat \mu} \delta \hat X^{\hat \nu} } \int_x \tilde e \, {\rm tr} \M &=
		\tilde e\left\{ 2 \hat \delta_{\hat \mu \hat \nu}\Delta
		- 2 g^{\mu\nu}\hat\delta_{\hat \mu \hat \nu}\,e^\alpha_{\hat\rho} \nabla_\mu \nabla_\alpha\hat X^{\hat\rho} \,\nabla_\nu
		\right.
		\\
		&\left. -4 g^{\mu\nu} \hat \delta_{\hat\rho(\hat\mu} e^\alpha_{\hat\nu)} \partial_\mu \hat X^{\hat\rho} \nabla_\nu \nabla_\alpha
		-4 g^{\mu\nu} \hat \delta_{\hat\rho(\hat\mu} e^\alpha_{\hat\nu)} \nabla_\mu \nabla_\alpha \hat X^{\hat\rho} \nabla_\nu
		\right\} \,, \nonumber
		\\
		\frac{1}{\sqrt{Z_N}}\frac{\delta^2}{\delta \hat X^{\hat \mu} \delta g_{\mu\nu} } \int_x \tilde e \, {\rm tr} \M &=
		\tilde e \mathds{1}^{\mu\nu \alpha \beta}\left\{ -2 \,\hat\delta_{\hat \mu \hat \nu} \partial_\alpha \hat X^{\hat\mu} \nabla_\beta \right.
		\\
		&\left. - \hat\delta_{\hat\nu\hat\rho} e^\rho_{\hat\mu} \partial_\alpha \hat X^{\hat\nu} \partial_\beta \hat X^{\hat\rho}\nabla_\rho
		-2  \hat\delta_{\hat\nu\hat\rho} e^\rho_{\hat\mu} \nabla_\alpha \nabla_\rho\hat X^{\hat\nu} \partial_\beta \hat X^{\hat\rho}
		+ 2  \hat\delta_{\hat\nu\hat\rho} e^\rho_{\hat\mu} \nabla_\alpha\nabla_\rho \hat X^{\hat\nu} \partial_{\beta} \hat X^{\hat\rho}
		\right\} \,,\nonumber
		\\
		\frac{1}{\sqrt{Z_N}}\frac{\delta^2}{\delta g_{\mu\nu} \delta \hat X^{\hat \mu} } \int_x \tilde e \, {\rm tr} \M &=
		\tilde e \mathds{1}^{\mu\nu \alpha \beta}\left\{ 2 \,\hat\delta_{\hat \mu \hat \nu} \partial_\alpha \hat X^{\hat\mu} \nabla_\beta
		+ 2 \,\hat\delta_{\hat \mu \hat \nu} \nabla_\alpha \nabla_\beta \hat X^{\hat\nu}
		+ 2 \,\hat\delta_{\hat \mu \hat \nu} e^\rho_{\hat\rho}\nabla_\alpha \nabla_\rho \hat X^{\hat\rho} \partial_\beta \hat X^{\hat\nu}
		\right.
		\\
		&\left. + \hat\delta_{\hat\nu\hat\rho} e^\rho_{\hat\mu} \partial_\alpha \hat X^{\hat\nu} \partial_\beta \hat X^{\hat\rho}\nabla_\rho+ 2  \hat\delta_{\hat\nu\hat\rho} e^\rho_{\hat\mu} \nabla_\alpha\nabla_\rho \hat X^{\hat\nu} \partial_{\beta} \hat X^{\hat\rho}
		\right\} \,. \nonumber
	\end{align}
	The Hessian of the term associated to $\alpha_R$ is
	\begin{align}
		\frac{1}{Z_N}g_{\rho\alpha}g_{\sigma\beta}\frac{\delta^2}{\delta g_{\mu\nu}\delta g_{\alpha\beta}}\int_x \tilde e \, R &=
		\tilde e\,\left\{
		-\frac{1}{2} \left( \mathds{1}^{\mu\nu}{}_{\rho\sigma}+g^{\mu\nu}g_{\rho\sigma}\right)\Delta
		- 2\delta^{(\mu}_{(\rho}g^{\nu)\alpha}\delta^{\beta}_{\sigma)} \nabla_{(\alpha}\nabla_{\beta)}
		+\delta^{(\mu}_{(\rho} \,R_{\sigma)}^{\nu)} +R^{(\mu}{}_{(\rho}{}^{\nu)}{}_{\sigma)}
		\right\} \\
		&+\tilde e e^\delta_{\hat \mu} \left(\nabla^\gamma \nabla_\delta \hat X^{\hat\mu} \right)
		\left\{
		-\frac{3}{2} \mathds{1}^{\mu\nu}{}_{\rho\sigma} \nabla_\gamma
		-\frac{1}{2} g^{\mu\nu}g_{\rho\sigma}\nabla_\gamma
		\right\} \nonumber
		\\
		&+\tilde e e^\delta_{\hat \mu} \left\{
		3 \delta^{(\mu}_{(\rho}g^{\nu)\alpha}\delta^{\beta}_{\sigma)}
		\left(\nabla_\delta\nabla_{(\alpha} \hat X^{\hat\mu}\right) \nabla_{\beta)}
		-\left(\mathds{1}^{\mu\nu\alpha\beta} g_{\rho\sigma} +
		\mathds{1}_{\rho\sigma}{}^{\alpha\beta} g^{\mu\nu} \right)\left(\nabla_\delta\nabla_{(\alpha} \hat X^{\hat\mu}\right) \nabla_{\beta)}
		\right\} \,,\nonumber
		\\
		\frac{\delta^2}{\delta \hat X^{\hat \mu} \delta \hat X^{\hat \nu} } \int_x \tilde e \, R &=
		0 \,,
		\\
		\frac{1}{\sqrt{Z_N}}\frac{\delta^2}{\delta \hat X^{\hat \mu} \delta g_{\mu\nu} } \int_x \tilde e \, R &=
		\tilde e\, e^\rho_{\hat\mu} \mathds{1}^{\mu\nu \alpha \beta}\left\{
		\nabla_\alpha \nabla_\beta + g_{\alpha\beta} \Delta - R_{\alpha\beta}
		\right\}\nabla_\rho \,,
		\\
		\frac{1}{\sqrt{Z_N}}\frac{\delta^2}{\delta g_{\mu\nu} \delta \hat X^{\hat \mu} } \int_x \tilde e \, R &=
		-\tilde e\, e^\rho_{\hat\mu} \mathds{1}^{\mu\nu \alpha \beta}\left\{
		\nabla_\rho \nabla_\alpha \nabla_\beta + g_{\alpha\beta} \nabla_\rho\Delta - R_{\alpha\beta}\nabla_\rho
		\right\} \,.
	\end{align}

	\section{Calculations of Section IV}\label{app:calculations}
	
	For the action in Equation~\eqref{eq:ansatzgamma}, in Appendix~\ref{sec:firstpossibility} we define the new covariant derivative in Equation~\eqref{eq:newcovder} in order to have the propagator in the form given in~\eqref{eq:proptot}.
	The same must be done also in the Hessian of the observable. Therefore, we change the ``free'' derivatives acting on the right with the new covariant derivative~\eqref{eq:newcovder}.
	
	Using the following identities of the Heat Kernel machinery \cite{Groh:2011dw}
	\bea
	H &=&\frac{1}{( 4\pi s)^{2} }(A_0 +s\,A_1)\,,\\
	H_\mu &=&\frac{1}{( 4\pi s)^{2} }(\mathcal{D}_\mu A_0 +s\,\mathcal{D}_\mu \,A_1)\,,\\
	H_{(\mu \nu)} (x,s) &=& \frac{1}{( 4\pi s)^{2} }\left(-\frac{1}{2s}g_{\mu \nu}A_0
	-\frac{1}{2}g_{\mu \nu}A_1+\mathcal{D}_{(\mu}\mathcal{D}_{\nu)}A_0 \right)\,,
	\eea
	where $s$ represents the proper time and
	\bea
	&A_0 =1 \,, \hspace{2cm}
	\mathcal{D}_\mu A_0 =0 \,,\hspace{2cm}
	\mathcal{D}_{(\mu}\mathcal{D}_{\nu)} A_0=\frac{1}{6}R_{\mu \nu} \,, \,\,\,\,\,\,\,\ldots\\
	&A_1 = \frac{1}{6}R \,,\hspace{2cm}
	\mathcal{D}_\mu A_1  =  -\frac{1}{2}\nabla_\mu E + \frac{1}{6}\nabla_\nu\Omega{}^\nu{}_{\mu} +\frac{1}{12}\nabla_\mu R \,, \,\,\,\,\,\,\ldots
	\eea
	it is clear that all the terms in the Hessian of the observables with one derivative acting on the right contribute to order $\partial^4$ since they are proportional to the curvature or to at least one derivative of the scalar fields. The terms with three derivatives contains as well higher order terms.

	Analysing in a schematic way the core of the rhs of Equation~\eqref{flow_co}, we have
	\begin{align}
		&{\rm Tr } \left[\mathcal{G}_k \cdot \left(\Gamma_k^{{\rm rel.}(2)} \right) \cdot \mathcal{G}_k \cdot\partial_t \mathcal{R}_k\right]=
		{\rm Tr } \left[ \frac{1}{{\rm det}g}
		\begin{pmatrix}
			\mathcal{G}_{gg} & 0  \\
			0 & \mathcal{G}_{XX}
		\end{pmatrix} \cdot
		\begin{pmatrix}
			\left(\Gamma_k^{{\rm rel.}(2)} \right)_{gg} & \left(\Gamma_k^{{\rm rel.}(2)} \right)_{gX}  \\
			\left(\Gamma_k^{{\rm rel.}(2)} \right)_{Xg} & \left(\Gamma_k^{{\rm rel.}(2)} \right)_{XX}
		\end{pmatrix}\cdot
		\begin{pmatrix}
			\mathcal{G}_{gg} \cdot\partial_t \mathcal{R}_k^{hh} & 0  \\
			0 & \mathcal{G}_{XX} \cdot\partial_t \mathcal{R}_k^{xx}
		\end{pmatrix}
		\right] \nn
		& = {\rm Tr} \left[ \frac{1}{{\rm det}g}
		\begin{pmatrix}
			\mathcal{G}_{gg}\cdot \left(\Gamma_k^{{\rm rel.}(2)} \right)_{gg}\cdot \mathcal{G}_{gg}\cdot\partial_t \mathcal{R}_k^{hh} & \mathcal{G}_{gg} \cdot\left(\Gamma_k^{{\rm rel.}(2)} \right)_{gX}\cdot \mathcal{G}_{XX} \cdot\partial_t \mathcal{R}_k^{xx}  \\
			\mathcal{G}_{XX} \cdot \left(\Gamma_k^{{\rm rel.}(2)} \right)_{Xg}\cdot\mathcal{G}_{gg}\cdot\partial_t \mathcal{R}_k^{hh} &\hspace{0.5cm} \mathcal{G}_{XX}\cdot\left(\Gamma_k^{{\rm rel.}(2)} \right)_{XX} \cdot \mathcal{G}_{XX}\cdot\partial_t \mathcal{R}_k^{xx}
		\end{pmatrix}
		\right]\nn
		&=  {\rm Tr} \left[ \frac{1}{\sqrt{{\rm det}g}} \mathcal{G}_{gg}\cdot \left(\Gamma_k^{{\rm rel.}(2)} \right)_{gg}\cdot \mathcal{G}_{gg} \cdot\partial_t \left(\frac{\mathcal{R}_k^{hh}}{\sqrt{{\rm det}g}} \right) \right]
		+  {\rm Tr} \left[ \frac{1}{\sqrt{{\rm det}g}} \mathcal{G}_{XX}\cdot\left(\Gamma_k^{{\rm rel.}(2)} \right)_{XX} \cdot \mathcal{G}_{XX}\cdot\partial_t \left(\frac{\mathcal{R}_k^{xx}}{\sqrt{{\rm det}g}} \right) \right]\,.
		\label{eq:traceobsscagra}
	\end{align}
	
	We can see that defining the covariant derivative~\eqref{eq:newcovder}, it is possible to simplify the analysis also for the flow of the observable since we don't need the mixed terms of the Hessian at second order in derivative.
	
	In Equation \eqref{eq:traceobsscagra} then the propagators must be expanded using Equations~\eqref{eq:propgrav} and \eqref{eq:propsca} keeping only terms at order $\partial^2$.
	
	Moreover, note that the first factor $1/\sqrt{{\rm det}g}$ simplify with the factor $\sqrt{{\rm det}g}$ coming from the heat kernel coefficients, while the second factor $1/\sqrt{{\rm det}g}$ under the regulators cancels using the Equation~\eqref{eq:regchoice}, which is needed to have the form~\eqref{eq:proptot} for the propagator. This implies that in all the integrals we find only the factor $\tilde e$ in the measure, like the rhs of~\eqref{eq:observablechoosen}.\\
	At this point, we are ready to calculate the rhs of \eqref{flow_co} using \eqref{eq:traceobsscagra}.\\
	
	The contributions on the rhs proportional to $\alpha_1$ from graviton
	\begin{align}
		= -\frac{\alpha_1\,G_N}{2\pi} \left(\int_x \tilde e \,8 \,{\rm tr}\,\M \right) \,
		Q_2\left[ \frac{\left(\partial_t -\eta_\phi \right) \mathcal{R}_k }{\left( P_k -2\Lambda \right)^2} \right]\,.
		\label{Q1}
	\end{align}
	The contributions on the rhs proportional to $\alpha_1$ from scalar
	\begin{align}
		&= -\frac{\alpha_1}{16 \pi^2} \left(\int_x \tilde e \,8 \right) \,
		Q_3\left[ \frac{\left(\partial_t -\hat\eta_k \right) \mathcal{R}_k }{ P_k^2} \right]
		-\frac{\alpha_1}{16 \pi^2} \left(\int_x \tilde e \,\frac{2}{3} R\right) \,
		Q_2\left[ \frac{\left(\partial_t -\hat\eta_k \right) \mathcal{R}_k }{ P_k^2} \right] \\
		& \hspace{1cm}- \frac{\alpha_1\, G_N}{\pi} \left(\int_x \tilde e \,\frac{5}{8} \,{\rm tr}\,\M \right)
		Q_3\left[ \frac{\left(\partial_t -\hat\eta_k \right) \mathcal{R}_k }{ P_k^3} \right] \,.
		\nonumber
		\label{Q2}
	\end{align}
	The contributions on the rhs proportional to $\alpha_R$ from graviton
	\begin{align}
		&= -\frac{\alpha_R\,G_N}{\pi} \left(\int_x \tilde e \,12  \right) \,
		Q_3\left[ \frac{\left(\partial_t -\eta_\phi \right) \mathcal{R}_k }{\left( P_k -2\Lambda \right)^2} \right]
		-\frac{\alpha_R\,G_N}{2\pi} \left(\int_x \tilde e \,\frac{10}{3} R  \right) \,
		Q_2\left[ \frac{\left(\partial_t -\eta_\phi \right) \mathcal{R}_k }{\left( P_k -2\Lambda \right)^2} \right]
		\\
		& \hspace{1cm} + \frac{\alpha_R\, G_N}{\pi} \int_x \tilde e\left( \,12 R -\frac{3}{8} 16\pi G_N\,{\rm tr}\,\M \right)
		Q_3\left[ \frac{\left(\partial_t -\eta_\phi \right) \mathcal{R}_k }{ (P_k-2\Lambda)^3} \right]\,. \nonumber
		\label{Q3}
	\end{align}
	The contributions on the rhs proportional to $\alpha_R$ from the  scalars are zero.\\
	Therefore, putting all the contributions together
	\begin{align}
		&\partial_t\alpha_0 = -\frac{\alpha_1}{2 \pi^2}
		Q_3\left[ \frac{\left(\partial_t -\hat\eta_k \right) \mathcal{R}_k }{ P_k^2} \right]
		-\frac{ 12\alpha_R\,G_N}{\pi}
		Q_3\left[ \frac{\left(\partial_t -\eta_\phi \right) \mathcal{R}_k }{\left( P_k -2\Lambda \right)^2} \right]\,,
		\\
		&\partial_t\alpha_R  -\gamma_g \, \alpha_R=
		-\frac{\alpha_1}{24 \pi^2}  \,
		Q_2\left[ \frac{\left(\partial_t -\hat\eta_k \right) \mathcal{R}_k }{ P_k^2} \right]
		-\frac{5\alpha_R\,G_N}{3\pi} \,
		Q_2\left[ \frac{\left(\partial_t -\eta_\phi \right) \mathcal{R}_k }{\left( P_k -2\Lambda \right)^2} \right]
		+ \frac{12\alpha_R\, G_N}{\pi}
		Q_3\left[ \frac{\left(\partial_t -\eta_\phi \right) \mathcal{R}_k }{ (P_k-2\Lambda)^3} \right]\,,
		\\
		&\partial_t\alpha_1-\left( \gamma_g + \hat\eta_k \right) \alpha_1=
		-\frac{4\alpha_1\,G_N}{\pi} \,
		Q_2\left[ \frac{\left(\partial_t -\eta_\phi \right) \mathcal{R}_k }{\left( P_k -2\Lambda \right)^2} \right]
		- \frac{5\alpha_1\, G_N}{8\pi}
		Q_3\left[ \frac{\left(\partial_t -\hat\eta_k \right) \mathcal{R}_k }{ P_k^3} \right]
		-6 \alpha_R\, G_N^2
		Q_3\left[ \frac{\left(\partial_t -\eta_\phi \right) \mathcal{R}_k }{ (P_k-2\Lambda)^3} \right]\,.
	\end{align}


\begin{thebibliography}{99}
		
		
		\bibitem{Weinberg}
		S. Weinberg,
		``Ultraviolet Divergences In Quantum Theories Of Gravitation,"
		In  \emph{General Relativity}; Hawking, S.W., Israel, W., Eds.; Cambridge University Press: Cambridge, UK, 1979; pp. 790--831.
		
		
		
		\bibitem{perbook}
		R. Percacci,
		\emph{An Introduction to Covariant Quantum Gravity and Asymptotic Safety};
		World Scientific: Singapore, 2017.
		
		\bibitem{rsbook}
		M. Reuter and F. Saueressig,
		\emph{Quantum Gravity and the Functional Renormalization Group: The Road towards Asymptotic Safety};
		CUP: Cambridge, UK, 2019.
		
		
		
		
		\bibitem{Donoghue:2019clr}
		J.~F.~Donoghue,
		``A Critique of the Asymptotic Safety Program,''
		Front. in Phys. \textbf{8} (2020), 56
		[arXiv:1911.02967 [hep-th]].
		\bibitem{Bonanno:2020bil}
		A.~Bonanno, A.~Eichhorn, H.~Gies, J.~M.~Pawlowski, R.~Percacci, M.~Reuter, F.~Saueressig and G.~P.~Vacca,
		``Critical reflections on asymptotically safe gravity,''
		Front. in Phys. \textbf{8} (2020), 269
		[arXiv:2004.06810 [gr-qc]].
		\bibitem{Knorr:2019atm}
		B.~Knorr, C.~Ripken and F.~Saueressig,
		``Form Factors in Asymptotic Safety: conceptual ideas and computational toolbox,''
		Class. Quant. Grav. \textbf{36} (2019) no.23, 234001
		[arXiv:1907.02903 [hep-th]].
		
		\bibitem{Pawlowski:2005xe}
		J.~M.~Pawlowski,
		``Aspects of the functional renormalisation group,''
		Annals Phys. \textbf{322} (2007), 2831-2915
		[arXiv:hep-th/0512261 [hep-th]].
		
		\bibitem{Herbst:2015ona}
		T.~K.~Herbst, J.~Luecker and J.~M.~Pawlowski,
		``Confinement order parameters and fluctuations,''
		[arXiv:1510.03830 [hep-ph]].
		
		
		
		\bibitem{Pagani:2016dof}
		C.~Pagani and M.~Reuter,
		``Composite Operators in Asymptotic Safety,''
		Phys. Rev. D \textbf{95} (2017) no.6, 066002
		[arXiv:1611.06522 [gr-qc]].
		
		
		
		\bibitem{Becker:2018quq}
		M.~Becker and C.~Pagani,
		``Geometric operators in the asymptotic safety scenario for quantum gravity,''
		Phys. Rev. D \textbf{99} (2019) no.6, 066002
		[arXiv:1810.11816 [gr-qc]].
		\bibitem{Becker:2019fhi}
		M.~Becker, C.~Pagani and O.~Zanusso,
		``Fractal Geometry of Higher Derivative Gravity,''
		Phys. Rev. Lett. \textbf{124} (2020) no.15, 151302
		[arXiv:1911.02415 [gr-qc]].
		
		\bibitem{Pagani:2016pad}
		C.~Pagani,
		``Note on scaling arguments in the effective average action formalism,''
		Phys. Rev. D \textbf{94} (2016) no.4, 045001
		[arXiv:1603.07250 [hep-th]].
		
		\bibitem{Pagani:2017tdr}
		C.~Pagani and H.~Sonoda,
		``Products of composite operators in the exact renormalization group formalism,''
		PTEP \textbf{2018} (2018) no.2, 023B02
		[arXiv:1707.09138 [hep-th]].
		\bibitem{Pagani:2020ejb}
		C.~Pagani and H.~Sonoda,
		``Operator product expansion coefficients in the exact renormalization group formalism,''
		Phys. Rev. D \textbf{101} (2020) no.10, 105007
		[arXiv:2001.07015 [hep-th]].
		
		\bibitem{Reuter:1996cp}
		M.~Reuter,
		``Nonperturbative evolution equation for quantum gravity,''
		Phys. Rev. D \textbf{57} (1998), 971-985
		[arXiv:hep-th/9605030 [hep-th]].
		
		\bibitem{Reuter:2001ag}
		M.~Reuter and F.~Saueressig,
		``Renormalization group flow of quantum gravity in the Einstein-Hilbert truncation,''
		Phys. Rev. D \textbf{65} (2002), 065016
		[arXiv:hep-th/0110054 [hep-th]].
		
		\bibitem{Souma:1999at}
		W.~Souma,
		``Nontrivial ultraviolet fixed point in quantum gravity,''
		Prog. Theor. Phys. \textbf{102} (1999), 181-195
		[arXiv:hep-th/9907027 [hep-th]].
		
		\bibitem{Falls:2014tra}
		K.~Falls, D.~F.~Litim, K.~Nikolakopoulos and C.~Rahmede,
		``Further evidence for asymptotic safety of quantum gravity,''
		Phys. Rev. D \textbf{93} (2016) no.10, 104022
		[arXiv:1410.4815 [hep-th]].
		
		\bibitem{Houthoff:2020zqy}
		W.~Houthoff, A.~Kurov and F.~Saueressig,
		``On the scaling of composite operators in asymptotic safety,''
		JHEP \textbf{04} (2020), 099
		[arXiv:2002.00256 [hep-th]].
		\bibitem{Kurov:2020csd}
		A.~Kurov and F.~Saueressig,
		``On characterizing the Quantum Geometry underlying Asymptotic Safety,''
		Front. in Phys. \textbf{8} (2020), 187
		[arXiv:2003.07454 [hep-th]].
		
		\bibitem{Rovelli:2001bz}
		C.~Rovelli,
		``Partial observables,''
		Phys. Rev. D \textbf{65} (2002), 124013
		[arXiv:gr-qc/0110035 [gr-qc]].
		\bibitem{Rovelli:1990ph}
		C.~Rovelli,
		``What Is Observable in Classical and Quantum Gravity?,''
		Class. Quant. Grav. \textbf{8} (1991), 297-316.
		\bibitem{Rovelli:1990pi}
		C.~Rovelli,
		``Quantum Reference Systems,''
		Class. Quant. Grav. \textbf{8} (1991), 317-332.
		\bibitem{Torre:1994ef}
		C.~G.~Torre,
		``The Problems of time and observables: Some recent mathematical results,''
		[arXiv:gr-qc/9404029 [gr-qc]].
		\bibitem{Torre:1993fq}
		C.~G.~Torre,
		``Gravitational observables and local symmetries,''
		Phys. Rev. D \textbf{48} (1993), R2373-R2376
		[arXiv:gr-qc/9306030 [gr-qc]].
		\bibitem{Dittrich:2005kc}
		B.~Dittrich,
		``Partial and complete observables for canonical general relativity,''
		Class. Quant. Grav. \textbf{23} (2006), 6155-6184
		[arXiv:gr-qc/0507106 [gr-qc]].
		\bibitem{Dittrich:2004cb}
		B.~Dittrich,
		``Partial and complete observables for Hamiltonian constrained systems,''
		Gen. Rel. Grav. \textbf{39} (2007), 1891-1927
		[arXiv:gr-qc/0411013 [gr-qc]].
		\bibitem{Dittrich:2015vfa}
		B.~Dittrich, P.~A.~Hoehn, T.~A.~Koslowski and M.~I.~Nelson,
		``Chaos, Dirac observables and constraint quantization,''
		[arXiv:1508.01947 [gr-qc]].
		\bibitem{Westman:2007yx}
		H.~Westman and S.~Sonego,
		``Coordinates, observables and symmetry in relativity,''
		Annals Phys. \textbf{324} (2009), 1585-1611
		[arXiv:0711.2651 [gr-qc]].
		\bibitem{Gielen:2018fqv}
		S.~Gielen,
		``Group field theory and its cosmology in a matter reference frame,''
		Universe \textbf{4} (2018) no.10, 103
		[arXiv:1808.10469 [gr-qc]].
		\bibitem{Donnelly:2016rvo}
		W.~Donnelly and S.~B.~Giddings,
		``Observables, gravitational dressing, and obstructions to locality and subsystems,''
		Phys. Rev. D \textbf{94} (2016) no.10, 104038
		[arXiv:1607.01025 [hep-th]].
		\bibitem{Giddings:2005id}
		S.~B.~Giddings, D.~Marolf and J.~B.~Hartle,
		``Observables in effective gravity,''
		Phys. Rev. D \textbf{74} (2006), 064018
		[arXiv:hep-th/0512200 [hep-th]].
		\bibitem{Hoehn:2020epv}
		P.~A.~Hoehn, A.~R.~H.~Smith and M.~P.~E.~Lock,
		``Equivalence of Approaches to Relational Quantum Dynamics in Relativistic Settings,''
		Front. in Phys. \textbf{9} (2021), 181
		[arXiv:2007.00580 [gr-qc]].
		\bibitem{DeFelice:2010uvx}
		F.~De Felice and D.~Bini,
		``Classical Measurements in Curved Space-Times,'' Cambridge University Press (2010)
		
		
		\bibitem{Komar:1958ymq}
		A.~Komar,
		``Construction of a Complete Set of Independent Observables in the General Theory of Relativity,''
		Phys. Rev. \textbf{111} (1958) no.4, 1182.
		
		\bibitem{Bergmann:1960wb}
		P.~G.~Bergmann and A.~B.~Komar,
		``Poisson brackets between locally defined observables in general relativity,''
		Phys. Rev. Lett. \textbf{4} (1960), 432-433.
		
		\bibitem{Bergmann:1961wa}
		P.~G.~Bergmann,
		``Observables in General Relativity,''
		Rev. Mod. Phys. \textbf{33} (1961), 510-514.
		
		\bibitem{Kuchar:1990vy}
		K.~V.~Kuchar and C.~G.~Torre,
		``Gaussian reference fluid and interpretation of quantum geometrodynamics,''
		Phys. Rev. D \textbf{43} (1991), 419-441.
		\bibitem{Brown:1994py}
		J.~D.~Brown and K.~V.~Kuchar,
		``Dust as a standard of space and time in canonical quantum gravity,''
		Phys. Rev. D \textbf{51} (1995), 5600-5629
		[arXiv:gr-qc/9409001 [gr-qc]].
		
		
		\bibitem{Khavkine:2015fwa}
		I.~Khavkine,
		``Local and gauge invariant observables in gravity,''
		Class. Quant. Grav. \textbf{32} (2015) no.18, 185019
		[arXiv:1503.03754 [gr-qc]].
		
		\bibitem{Brunetti:2013maa}
		R.~Brunetti, K.~Fredenhagen and K.~Rejzner,
		``Quantum gravity from the point of view of locally covariant quantum field theory,''
		Commun. Math. Phys. \textbf{345} (2016) no.3, 741-779
		[arXiv:1306.1058 [math-ph]].
		
		\bibitem{Brunetti:2016hgw}
		R.~Brunetti, K.~Fredenhagen, T.~P.~Hack, N.~Pinamonti and K.~Rejzner,
		``Cosmological perturbation theory and quantum gravity,''
		JHEP \textbf{08} (2016), 032
		[arXiv:1605.02573 [gr-qc]].
		
		\bibitem{Frob:2017gyj}
		M.~B.~Fr\"ob and W.~C.~C.~Lima,
		``Propagators for gauge-invariant observables in cosmology,''
		Class. Quant. Grav. \textbf{35} (2018) no.9, 095010
		[arXiv:1711.08470 [gr-qc]].
		
		
		\bibitem{Tambornino:2011vg}
		J.~Tambornino,
		``Relational Observables in Gravity: a Review,''
		SIGMA \textbf{8} (2012), 017
		[arXiv:1109.0740 [gr-qc]].
		\bibitem{Ferrero:2020jts}
		R.~Ferrero and R.~Percacci,
		``Dynamical diffeomorphisms,''
		Class. Quant. Grav. \textbf{38} (2021) no.11, 115011
		[arXiv:2012.04507 [gr-qc]].
		\bibitem{Baldazzi:2021ydj}
		A.~Baldazzi, R.~B.~A.~Zinati and K.~Falls,
		``Essential renormalisation group,''
		[arXiv:2105.11482 [hep-th]].
		\bibitem{Baldazzi:2021orb}
		A.~Baldazzi and K.~Falls,
		``Essential Quantum Einstein Gravity,''
		[arXiv:2107.00671 [hep-th]].
		
		
		
		
		
		\bibitem{Wetterich:1992yh}
		C.~Wetterich,
		``Exact evolution equation for the effective potential,''
		Phys. Lett. B \textbf{301} (1993), 90-94
		[arXiv:1710.05815 [hep-th]].
		\bibitem{Morris:1993qb}
		T.~R.~Morris,
		``The Exact renormalization group and approximate solutions,''
		Int. J. Mod. Phys. A \textbf{9} (1994), 2411-2450
		[arXiv:hep-ph/9308265 [hep-ph]].
		
		
		
		
		\bibitem{Percacci:2003jz}
		R.~Percacci and D.~Perini,
		``Asymptotic safety of gravity coupled to matter,''
		Phys. Rev. D \textbf{68} (2003), 044018
		[arXiv:hep-th/0304222 [hep-th]].
		
		\bibitem{Laporte:2021kyp}
		C.~Laporte, A.~D.~Pereira, F.~Saueressig and J.~Wang,
		``Scalar-Tensor theories within Asymptotic Safety,''
		[arXiv:2110.09566 [hep-th]].
		\bibitem{Dona:2013qba}
		P.~Don\`a, A.~Eichhorn and R.~Percacci,
		``Matter matters in asymptotically safe quantum gravity,''
		Phys. Rev. D \textbf{89} (2014) no.8, 084035
		[arXiv:1311.2898 [hep-th]].
		
		\bibitem{Peli:2020yiz}
		Z.~P\'eli,
		``Derivative expansion for computing critical exponents of $O(N)$ symmetric models at next-to-next-to-leading order,''
		Phys. Rev. E \textbf{103} (2021) no.3, 032135
		[arXiv:2010.04020 [hep-th]].
		
		
		
		
		\bibitem{Groh:2011dw}
		K.~Groh, F.~Saueressig and O.~Zanusso,
		``Off-diagonal heat-kernel expansion and its application to fields with differential constraints,''
		[arXiv:1112.4856 [math-ph]].
		\bibitem{Steinwachs:2011zs}
		C.~F.~Steinwachs and A.~Y.~Kamenshchik,
		``One-loop divergences for gravity non-minimally coupled to a multiplet of scalar fields: calculation in the Jordan frame. I. The main results,''
		Phys. Rev. D \textbf{84} (2011), 024026
		[arXiv:1101.5047 [gr-qc]].
		
		
		\bibitem{Loll:2019rdj}
		R.~Loll,
		``Quantum Gravity from Causal Dynamical Triangulations: A Review,''
		Class. Quant. Grav. \textbf{37} (2020) no.1, 013002
		[arXiv:1905.08669 [hep-th]].
		\bibitem{Brunekreef:2020bwj}
		J.~Brunekreef and R.~Loll,
		``Curvature profiles for quantum gravity,''
		Phys. Rev. D \textbf{103} (2021) no.2, 026019
		[arXiv:2011.10168 [gr-qc]].
		\bibitem{Ambjorn:2021uge}
		J.~Ambj\o{}rn, Z.~Drogosz, J.~Gizbert-Studnicki, A.~G\"orlich, J.~Jurkiewicz and D.~N\'emeth,
		``Scalar fields in Causal Dynamical Triangulations,''
		[arXiv:2105.10086 [gr-qc]].
	\end{thebibliography}
\end{document}